\documentclass[12pt]{article}
\usepackage{amscd}
\usepackage{bbm}
\usepackage{mathrsfs}
\usepackage{amssymb}
\usepackage{graphics}
\usepackage{graphicx}
\usepackage{amsfonts}
\usepackage{amsmath}
\usepackage[numbers,sort&compress]{natbib}
\usepackage{multirow}
\usepackage{array}
\usepackage{float}
\usepackage{caption}
\usepackage[english]{babel}
\usepackage{hyphenat}
\usepackage{microtype}

\usepackage{titlesec}
\titleformat{\section}
  {\normalfont\Large\bfseries}{\thesection}{0.5em}{}
\titleformat{\subsection}
  {\normalfont\large\bfseries}{\thesubsection}{0.5em}{}

\begin{document}
\date{}
\title{An adaptive weighted extrapolation neural network for dispersive shock waves generated by square-root initial condition of the KdV equation}
\author{Xuan-Jie Wang, Rui Guo\thanks{Corresponding author, gr81@sina.com}, An-Yao Jin, Hua-Ying Ren \\
{\em
School of Mathematics, Taiyuan University  of} \\
{\em Technology, Taiyuan 030024, China} } \maketitle

\begin{abstract}

Simulating dispersive shock waves (DSWs) using physics-informed neural networks (PINNs) presents significant numerical difficulties. In particular, the square-root initial condition exhibits singularity near the origin, making it difficult to fit accurately. This initial fitting error is amplified through nonlinear evolution and gradually accumulates during extrapolation, imposing high demands on the long-term extrapolation capability of numerical methods. Accordingly, this paper proposes an improved physics-guided multistage neural network (IPgMSNN) framework. IPgMSNN is built upon the original PgMSNN framework, with its core innovation focused on the third-stage training strategy. Specifically, online fine-tuning and a weighted target mechanism are introduced, which effectively suppress error accumulation during long-term extrapolation. Numerical experiments systematically evaluate IPgMSNN from three perspectives: forward problem solving, model stability, and parameter inversion. Experimental results demonstrate that IPgMSNN effectively captures high-frequency wavefront details, maintains long-term extrapolation stability, and achieves high-precision parameter inversion, providing an efficient solution for deep learning-based simulation of dispersive nonlinear systems.

\vspace{7mm}\noindent\emph{Keywords}: Dispersive shock waves; Physics-informed neural networks; Square-root initial condition; Online fine-tuning; Weighted target mechanism
\end{abstract}

\newpage
\section{Introduction}
\hspace{1.5em}Physics-informed neural networks (PINNs) have emerged in recent years as a powerful class of numerical tools for solving nonlinear partial differential equations (PDEs)~\cite{raissi2019,kaltsas2025,alves2026,tartakovsky2024,castelano2024}. Unlike conventional methods such as finite difference and finite element schemes, PINNs bypass the need for mesh generation by adding the residuals of the governing equations, initial conditions, and boundary conditions as penalty terms into the loss function, and construct approximate solutions over a continuous spatiotemporal domain via automatic differentiation~\cite{shin2023,Dlin2022,karumuri2020}. In the field of nonlinear waves, the earliest successes of PINNs were achieved in the numerical reproduction of localized wave structures—solitons, breathers, and rogue waves. The three most representative classes of localized modes, have each been accurately captured by PINNs and their variants across a range of integrable systems~\cite{Slin2022,saharia2025,wang2021,liu2023,fang2021}. These studies have amply demonstrated the potential of PINNs for simulating nonlinear wave phenomena. However, they have largely focused on localized structures whose initial conditions are smooth and whose temporal evolution remains relatively regular. When the solution possesses rich multi-scale oscillatory features and its detailed waveform morphology cannot be directly inferred from the initial condition alone, how to maintain the accuracy and stability of PINNs becomes a critical challenge that remains to be resolved.

Different from the aforementioned localized waves, dispersive shock waves (DSWs) represent a typical class of nonlocal nonlinear wave structures~\cite{whitham2011,hoefer2009,el2016}. Their core characteristic lies in the fact that the nonlinear effects cause the wave front steepen and further result in breaking~\cite{ivanov2020b,gentilini2015}, while the dispersive effect regularizes it into an oscillatory train with decreasing amplitudes and gradually varying wavelengths. The dynamic balance between these two effects endows DSWs with extremely rich oscillatory structures. In the context of initial-value problems, DSWs are typically excited by two types of initial disturbances: one is discontinuous step initial conditions, upon which many classical theoretical works on DSWs are established~\cite{gurevich1973}; the other is continuous square-root-type initial conditions, which can naturally excite a complete DSWs oscillatory structure near the origin~\cite{kamchatnov2021,kamchatnov2018}. 

Our team has previously addressed the step initial value problem for the generalized Gardner equation by proposing a physics-guided multistage neural network (PgMSNN), which successfully achieved numerical simulation of DSWs across multiple parameter regions through the introduction of a dispersion factor and a Deep Runge-Kutta method~\cite{yuan2024}. However, directly extending this framework to square-root initial condition encounters a new difficulty. That is, the square-root initial condition exhibits singularity near the origin, with its gradient tending to infinity, which poses a higher demand on the network's ability to fit the initial condition. As a result, the original temporal recurrence strategy tends to accumulate phase errors during long-term extrapolation, causing the gradual loss of wavefront details. To address these difficulties, this paper introduces targeted improvements to the original framework. In particular, a weighted target mechanism and an online fine-tuning strategy are incorporated into the third stage, enabling the model to smoothly transition from relying on the exact solution to relying on physical evolution during extrapolation, thereby effectively suppressing error accumulation in long-term extrapolation and successfully achieving stable simulation of DSWs under square-root initial condition.

As a classical model for describing the evolution of DSWs in shallow water, the Korteweg-de Vries (KdV) equation occupies a central position in fluid mechanics~\cite{congy2019,hammack1974,crabb2021,ankiewicz2019}. Through the concise coupling of a nonlinear advection term and a third-order dispersive term, this equation accurately captures the dynamic balance between nonlinear steepening and dispersive spreading, thus providing an ideal theoretical framework for investigating DSWs dynamics. The KdV equation considered in this paper takes the form
\begin{equation}
u_t + 6uu_x + u_{xxx} = 0,
\label{eq:KdV equation}
\end{equation}
where $u = u(x,t)$ is a real-valued function of the spatial variable $x$ and time $t$, with subscripts denoting partial derivatives. Specifically, $u_t$, $u_x$, and $u_{xxx}$ represent the first-order time derivative, the first-order spatial derivative, and the third-order spatial derivative of $u(x,t)$, respectively. The initial condition is chosen as the square-root form
\begin{equation}
u_0(x) = u(x,0) =
\begin{cases}
\sqrt{-x}, & x < 0, \\
0, & x > 0.
\end{cases}
\label{eq:square-root form}
\end{equation}
This initial profile forms a continuous but singular transition region near the origin, which can naturally excite a complete DSWs oscillatory structure during the early stage of evolution. The wavefront exhibits a train of oscillatory waves with decreasing amplitude and varying wavelength. By means of Whitham modulation theory, the reference solution to this problem can be exactly constructed through Riemann invariants and elliptic functions, providing a rigorous benchmark for validating the reliability of numerical methods~\cite{kamchatnov2000}. Applying the improved physics-guided multistage neural network (IPgMSNN) model proposed above to this initial-value problem, the model not only accurately reproduces the phase and amplitude of the high-frequency oscillations at the wavefront, but also effectively suppresses error accumulation during long-term extrapolation, achieving satisfactory simulation results.

The remainder of this paper is organized as follows. Section 2 provides a brief review of the fundamental framework of the standard physics-informed neural network (Std-PINN). Section 3 elaborates on the proposed IPgMSNN model, sequentially introducing the dispersion factor mechanism, the learnable Runge-Kutta method, and the three-stage progressive training strategy. Section 4 derives the exact solution of the DSWs for the KdV equation under square-root initial condition via Whitham modulation theory, establishing a reference benchmark for subsequent experiments. Section 5 systematically compares the performance of Std-PINN, PgMSNN, and IPgMSNN in forward problem solving. Sections 6 and 7 discuss the numerical stability of all models and their applicability to inverse problems, respectively. Finally, Section 8 concludes the paper with a summary of the main contributions and outlines directions for future research.

\section{The Standard Physics Informed Neural Network for the KdV Equation}
\hspace{1.5em}PINNs approximate the nonlinear mapping between the solution of PDEs and their spatiotemporal coordinates by embedding physical laws into the loss function. As a fundamental variant of PINNs, the standard physics-informed neural network (Std-PINN) leverages automatic differentiation to compute derivatives of field variables and integrates initial and boundary value constraints into the training process. To achieve better simulation performance, various improvements have been made to the PINN model~\cite{perez2023,duarte2025,arzani2023,lantigua2026,grubas2023,amini2023}. In this study, Std-PINN is applied to solve the one-dimensional KdV equation with a square-root initial condition, aiming to investigate its performance in solving strongly nonlinear PDEs involving high-order derivatives.

The constructed Std-PINN takes spatiotemporal coordinates \((x, t)\) as inputs and the solution \(u(x, t)\) of the KdV equation as the output. It adopts a multi-layer perceptron (MLP) with a 2-50×7-1 architecture, where the hidden layers employ the sine function as the activation function to adapt to the oscillatory characteristics and nonlinear evolution laws of the KdV equation's solution~\cite{zhou2026}. By virtue of the automatic differentiation mechanism, higher-order derivatives of the network output are computed to construct the physical residual \(R:=\hat{u}_{t}+6 \hat{u} \hat{u}_{x}+\hat{u}_{xxx}\) (where \(\hat{u}(x, t)\) denotes the predicted output of the network). The embedding of physical constraints is achieved by minimizing this residual.

The loss function of the model consists of initial condition loss, data fitting loss, and PDE residual loss, all formulated in the form of mean squared error (MSE). The total loss function is a weighted sum of these three components:
\begin{equation}
L_{total}=\omega _{i}MSE_{I}+\omega _{d}MSE_{D}+\omega _{r}MSE_{R},
\end{equation}
where \(\omega_{i}\), \(\omega_{d}\) and \(\omega_{r}\) represent the weights of the respective loss terms. Each loss term is specifically defined as follows.

$\bullet$ Mean squared error on the initial condition
\begin{equation}
MSE_{I}=\frac{1}{N_{I}} \sum_{k=1}^{N_{I}}\left(\hat{u}\left(x_{k}^{I}, 0\right)-\Phi\left(x_{k}^{I}\right)\right)^{2},
\end{equation}
where \(N_{I}\) is the number of initial spatial sampling points, and \(x_{k}^{I}\) denotes the initial spatial sampling points.

$\bullet$ Mean squared error on the data fitting
\begin{equation}
MSE_{D}=\frac{1}{N_{D}} \sum_{k=1}^{N_{D}}\left(\hat{u}\left(x_{k}^{D}, t_{k}^{D}\right)-u_{exact }\left(x_{k}^{D}, t_{k}^{D}\right)\right)^{2},
\end{equation}
where \(N_{D}\) is the total number of spatiotemporal sampling points in the training segment, and \(u_{exact }(x_{k}^{D}, t_{k}^{D})\) is the exact value of the reference solution at the corresponding point.

$\bullet$ Mean squared error on the governing equation
\begin{equation}
MSE_{R}=\frac{1}{N_{R}} \sum_{k=1}^{N_{R}}\left(R\left(x_{k}^{R}, t_{k}^{R}\right)\right)^{2},
\end{equation}
where \(N_{R}\) is the number of residual collocation points, and \((x_{k}^{R}, t_{k}^{R})\) are randomly selected spatiotemporal sampling points in the entire domain.

\begin{figure}[t!]
\centering
\includegraphics[scale=0.4]{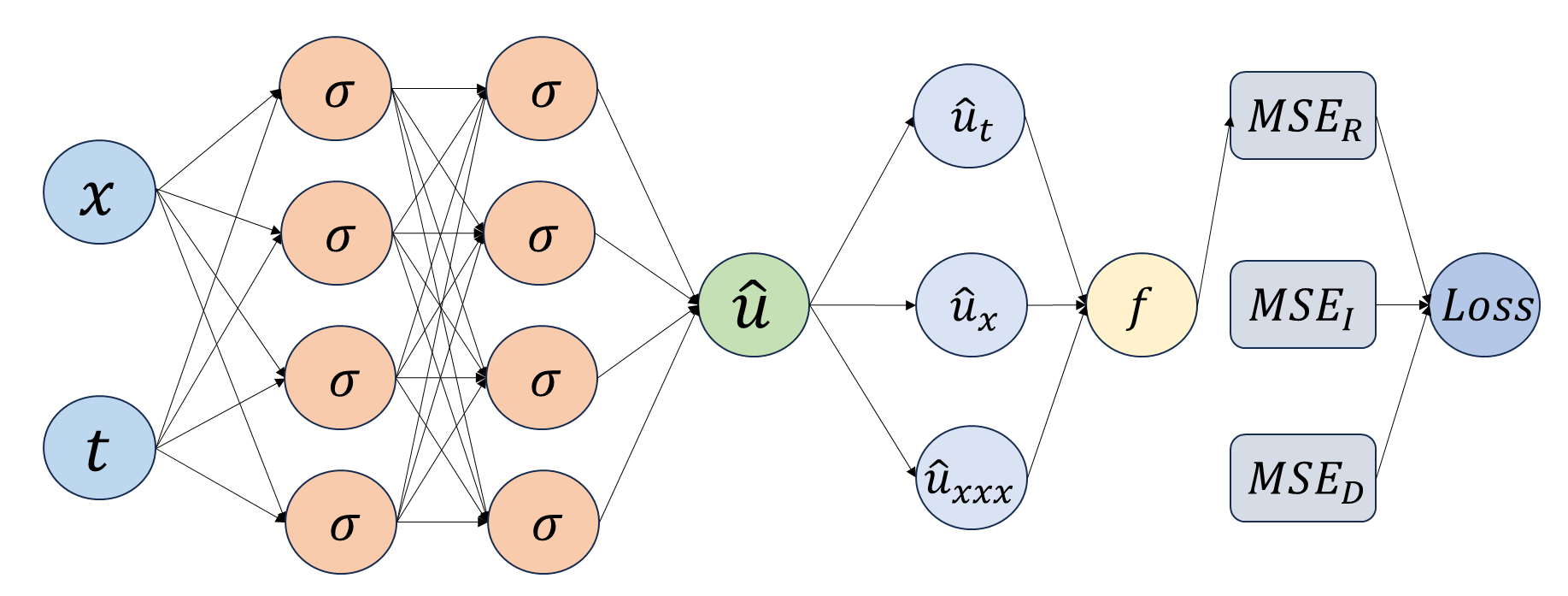}

\vspace{-0.2cm}
{\footnotesize \textbf{Fig.~1.} Std-PINN structure.}
\end{figure}

The Adam optimizer (with an initial learning rate of 0.001) is utilized for training, combined with a learning rate scheduler and a gradient clipping strategy (maximum norm of 1.0) to ensure training stability. A maximum of 3000 iterations and an early stopping strategy with a patience of 1000 are set to avoid overfitting. The time domain is discretized into 1000 time steps, which are divided into a training segment and an extrapolation segment at a ratio of 9:1, serving for model training and generalization ability verification, respectively.

Relative total error, MSE, root mean squared error (RMSE), and mean absolute error (MAE) are adopted as evaluation metrics. By comparing with the high-precision reference solution, the solution accuracy of the model in the training segment, extrapolation segment, and the entire domain is quantitatively evaluated, comprehensively reflecting the performance of Std-PINN in solving the KdV equation with a square-root initial condition. The model structure of the Std-PINN method is shown in Fig.~1.

\section{Improved Physics-guided Multistage Neural Network Method}
\hspace{1.5em}The proposed IPgMSNN framework comprises two complementary neural components—a Dispersion Physics-Informed Neural Network (DPINN) and a Deep Runge-Kutta Network (DRKT)—orchestrated through a three-stage progressive training strategy. The DPINN serves as the primary spatial approximator, responsible for reconstructing the wave packet profile at individual time instances with high fidelity, while the DRKT functions as a temporal evolution operator, learning the dynamical mapping between adjacent time steps. The core philosophy of the three-stage training strategy is as follows: first, enable the DPINN to fully learn the spatial structure of the initial wave packet; second, allow the DRKT to master the dynamical laws of temporal evolution; and finally, through an online fine-tuning mechanism during long-term extrapolation, facilitate the synergistic operation of both components, thereby effectively suppressing error accumulation while maintaining physical consistency.

\subsection{Dispersion Physics-Informed Neural Network Method}
\hspace{1.5em}When confronted with wave problems dominated by strong dispersive effects, original PINNs often struggle to adequately perceive high-frequency components, leading to excessive smoothing of wavefront details. To address this limitation, this paper introduces a trainable dispersion factor into the original PINN architecture, serving as a global scaling coefficient for the network output. This factor is updated synchronously with the network weights during training and can adaptively adjust the model's sensitivity to high-frequency oscillatory components, thereby significantly enhancing the network's capacity to represent complex wavefront structures.

Specifically, consider a fully connected feedforward network with $D+1$ layers, where the $d$-th layer ($d = 1,2,\dots,D$) contains $n_d$ neurons, and the sine activation function $\sin(\cdot)$ is uniformly adopted to accommodate the oscillatory characteristics of solutions to the KdV equation. Let the network input be the spatiotemporal coordinates $\mathbf{z} = (x, t)^{\intercal}$. The forward propagation process can be described as:
\begin{equation}
\mathbf{h}^{(1)} = \sin\left(\mathbf{W}^{(1)}\mathbf{z} + \mathbf{b}^{(1)}\right),
\end{equation}
\begin{equation}
\mathbf{h}^{(d)} = \sin\left(\mathbf{W}^{(d)}\mathbf{h}^{(d-1)} + \mathbf{b}^{(d)}\right), \quad d = 2,3,\dots, D-1,
\end{equation}
\begin{equation}
\mathbf{h}^{(D)} = \mathbf{W}^{(D)}\mathbf{h}^{(D-1)} + \mathbf{b}^{(D)},
\end{equation}
where $\mathbf{W}^{(d)} \in \mathbb{R}^{n_d \times n_{d-1}}$ and $\mathbf{b}^{(d)} \in \mathbb{R}^{n_d}$ denote the weight matrix and bias vector of the $d$-th layer, respectively. Based on this, a trainable dispersion factor $\alpha_{DF} \in \mathbb{R}^{+}$ is introduced and applied to the output layer, yielding the final prediction:
\begin{equation}
\hat{u}(\mathbf{z}; \Theta, \alpha_{DF}) = \alpha_{DF} \cdot \mathbf{h}^{(D)},
\end{equation}
where $\Theta = \{\mathbf{W}^{(d)}, \mathbf{b}^{(d)}\}_{d=1}^{D}$ denotes the collection of the network's base parameters. The dispersion factor $\alpha_{DF}$ is incorporated into the set of trainable parameters and jointly optimized with $\Theta$ via the backpropagation algorithm. Fig.~2 illustrates the model structure of the DPINN method, where the input layer receives the spatiotemporal coordinates $(x,t)$, the hidden layers apply successive sine-activated transformations, and the output is multiplied by the trainable dispersion factor $\alpha_{DF}$ to produce the predicted solution $\hat{u}(x,t)$.

\begin{figure}[t!]
\centering
\includegraphics[scale=0.4]{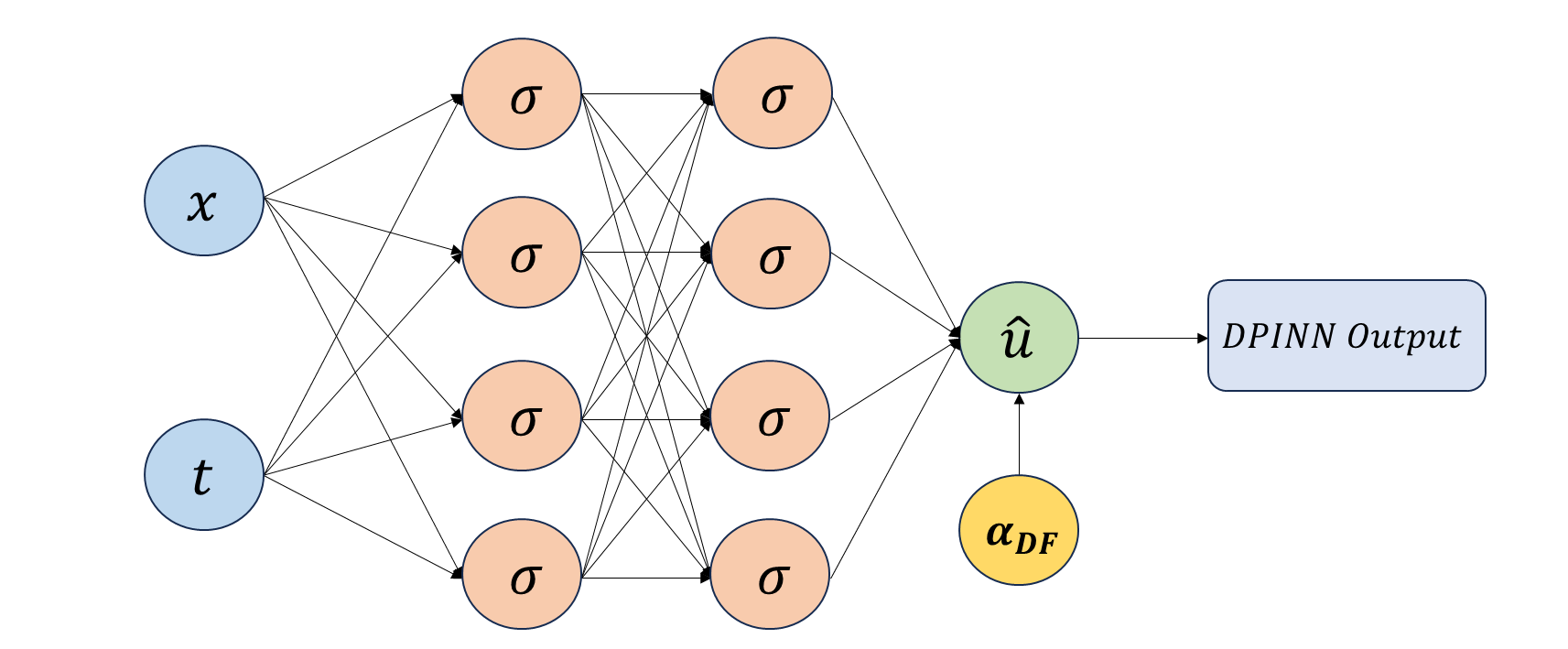}

\vspace{-0.2cm}
{\footnotesize \textbf{Fig.~2.} DPINN structure.}
\end{figure}

From a physical perspective, the dispersion factor $\alpha_{DF}$ functions as a frequency-dependent gain adjustment applied to the network output: When intense high-frequency oscillations are present near the wavefront, the optimization process adaptively tunes the value of $\alpha_{DF}$, enabling the network to respond more sensitively to abrupt variations in that region and thereby avoiding the loss of critical oscillatory details due to smoothing effects. Experimental results demonstrate that the introduction of the dispersion factor significantly enhances the model's ability to capture high-frequency oscillatory components in dispersion-dominated systems, providing high-quality initial state representations for the subsequent multi-stage training.

\subsection{Deep Runge-Kutta Method}
\hspace{1.5em}The Runge-Kutta method, as a classical algorithm for the numerical solution of ordinary differential equations, occupies a prominent position in scientific computing due to its high accuracy and favorable numerical stability. Addressing the temporal evolution problem of dispersive shock waves that is the focus of this paper, we propose a numerical scheme that deeply integrates deep learning mechanisms with the traditional Runge-Kutta framework, termed the DRKT Method. Unlike traditional approaches where the integration coefficients are predetermined through Taylor series expansions, DRKT treats the coefficients $\alpha_1, \alpha_2, \dots, \alpha_n$ as trainable parameters, adaptively optimizing them through the backpropagation mechanism during training, thereby enabling the time-stepping strategy to flexibly match the dynamical characteristics of the specific problem.

Rewrite the KdV equation in the evolutionary form:
\begin{equation}
\frac{\partial u}{\partial t} = \mathcal{F}[u], \quad \mathcal{F}[u] := -\left(6u\frac{\partial u}{\partial x} + \frac{\partial^3 u}{\partial x^3}\right),
\end{equation}
where $\mathcal{F}[\cdot]$ denotes the spatial differential operator. Let $\delta t$ be the time step size, and let $u^{(k)}$ represent the numerical solution at the $k$-th time level, with $u_x^{(k)}$ and $u_{xxx}^{(k)}$ denoting its first- and third-order spatial derivatives, respectively (computed via automatic differentiation). The recursive procedure of DRKT in the time direction can be expressed as:
\begin{equation}
u^{(k+1)} = u^{(k)} + \delta t \sum_{j=1}^{n} \alpha_j K_j,
\end{equation}
where the intermediate increments $K_j$ are computed sequentially as follows:
\begin{equation}
\left\{
\begin{aligned}
K_1 &= \mathcal{F}\left[u^{(k)}\right], \\
K_2 &= \mathcal{F}\left[u^{(k)} + \delta t \cdot K_1\right], \\
K_3 &= \mathcal{F}\left[u^{(k)} + \delta t \cdot K_2\right], \\
&\vdots \\
K_n &= \mathcal{F}\left[u^{(k)} + \delta t \cdot K_{n-1}\right].
\end{aligned}
\right.
\end{equation}

\begin{figure}[t!]
\centering
\includegraphics[scale=0.45]{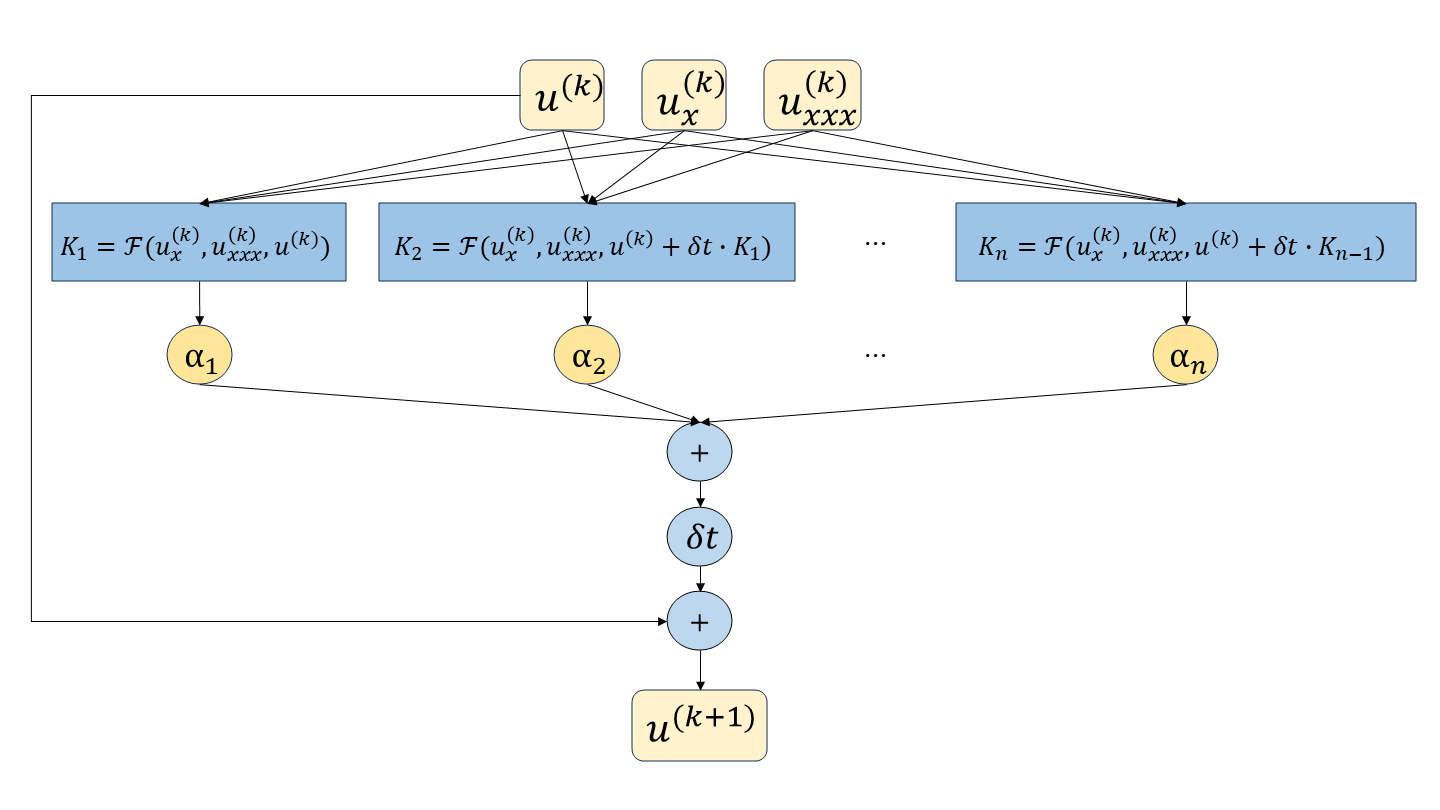}

\vspace{-0.2cm}
{\footnotesize \textbf{Fig.~3.} DRKT structure.}
\end{figure}

For brevity, the explicit dependence of $\mathcal{F}[\cdot]$ on the spatial derivatives $u_x^{(k)}$ and $u_{xxx}^{(k)}$ is omitted in the above expressions. The introduction of trainable parameters $\boldsymbol{\alpha} = (\alpha_1, \alpha_2, \dots, \alpha_n)^{\intercal}$ enables DRKT to automatically learn the optimal combination of integration coefficients from the training data, rather than being constrained by the fixed values imposed by accuracy requirements in traditional Runge-Kutta methods. Fig.~3 illustrates the model structure of the DRKT method, depicting the complete workflow: starting from the current state $u^{(k)}$, the intermediate increments $K_1, K_2, \dots, K_n$ are computed sequentially, weighted by the trainable coefficients $\alpha_j$, summed together, and finally added to $u^{(k)}$ to produce the state $u^{(k+1)}$ at the next time step.

The merit of this method is twofold. On the one hand, it inherits the stability advantages of the traditional Runge-Kutta framework in handling temporal evolution problems. On the other hand, the introduction of learnable coefficients $\boldsymbol{\alpha}$ endows the model with the flexibility to adaptively adjust its evolution strategy according to the specific problem at hand. It is worth emphasizing that what DRKT learns during this stage is essentially the dynamical evolution law governed by the KdV equation, rather than a mechanical fitting to the training data, which provides a theoretical guarantee for its generalization capability in unknown temporal regions.

\subsection{Three-Stage Progressive Training Strategy}
\hspace{1.5em}A single network architecture often struggles to simultaneously achieve high-precision reconstruction of the initial state and maintain dynamical consistency throughout the entire evolution process. To address this issue, this paper designs a three-stage progressive training strategy that organically integrates the static spatial representation capability of DPINN with the dynamic temporal evolution capability of DRKT, achieving a smooth transition from accurate fitting of the initial wave packet to stable long-term extrapolation. Fig.~4 presents the overall network architecture of the IPgMSNN model, clearly illustrating the synergistic relationship and data flow between the DPINN and DRKT components across the three stages.

\textbf{Stage 1: Initial Wave Packet Reconstruction.} This stage focuses on simulating the initial generation process of the DSWs. The loss function consists solely of a data fidelity term without imposing physical residual constraints, and is formulated as:
\begin{equation}
\mathcal{L}_1 = \frac{1}{N_u}\sum_{i=1}^{N_u} \left| \hat{u}_{\text{DPINN}}^{(i)}(\mathbf{z}_0; \Theta, \alpha_{DF}) - u_E^{(i)} \right|,
\end{equation}
where $N_u$ denotes the total number of sampling points, $u_E$ represents the exact reference solution, and $\mathbf{z}_0 = (x, 0)^{\intercal}$ corresponds to the spatiotemporal coordinates at the initial instant. Adopting the mean absolute error in place of the mean squared error as the loss metric effectively avoids numerical anomalies caused by gradient explosion. The initial wave packet excited by the square-root initial condition contains the key spectral information required for subsequent evolution; prioritizing fitting accuracy at this stage helps establish a reliable foundation for the temporal advancement that follows.

\begin{figure}[t!]
\centering
\includegraphics[scale=0.4]{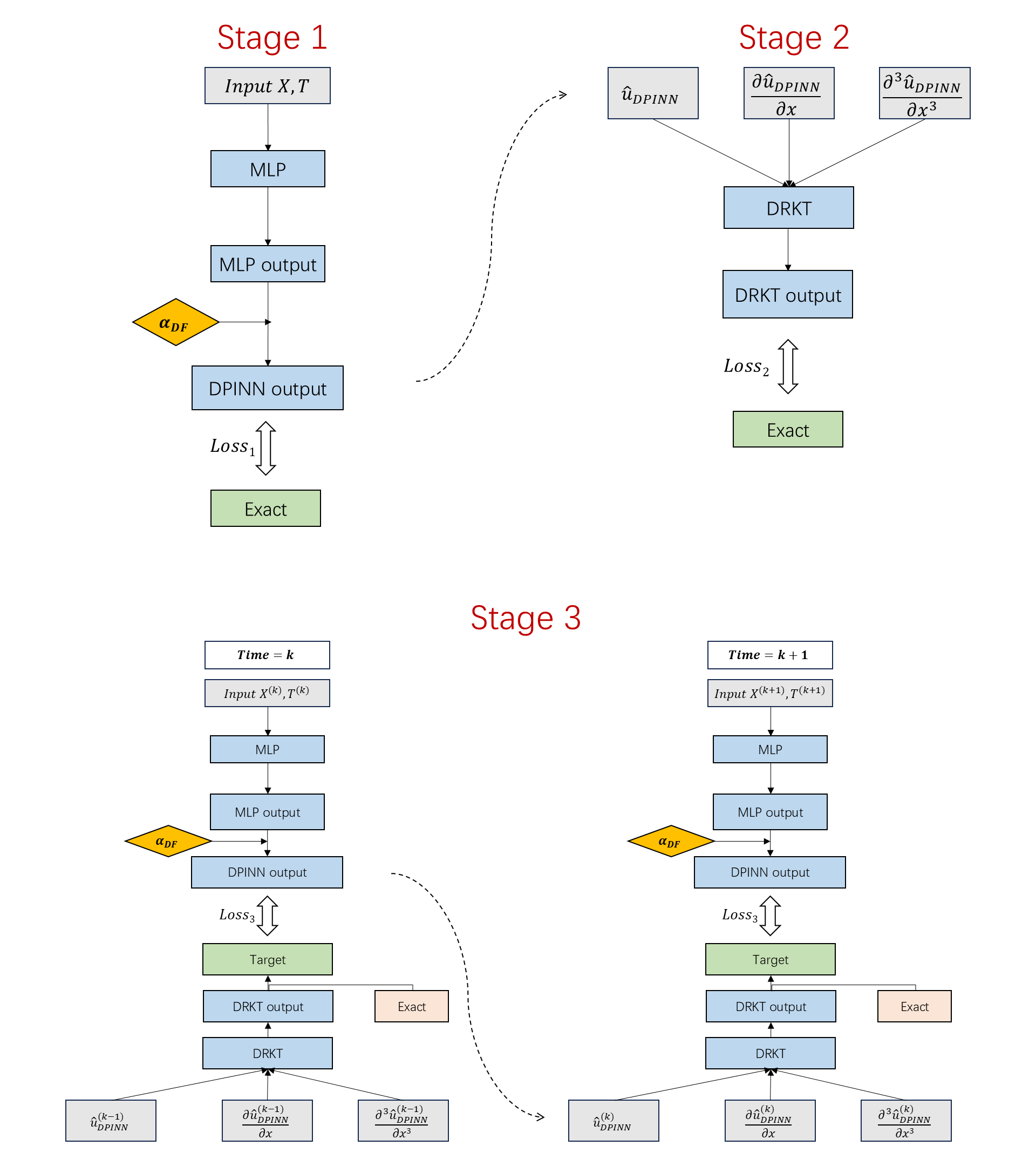}
\vspace{-0.2cm}

{\footnotesize \textbf{Fig.~4.} IPgMSNN structure.}
\end{figure}

\textbf{Stage 2: Evolution Dynamics Learning.} This stage employs the converged output of DPINN from Stage~1 as benchmark data to drive the DRKT network to learn the evolutionary mapping between adjacent time steps. Specifically, the solution function $\hat{u}_{\text{DPINN}}$ obtained in Stage~1, along with its spatial derivatives $\hat{u}_x$ and $\hat{u}_{xxx}$, is used as input to train the DRKT's ability to predict the state at the next time step from the current state. The loss function is defined as the MAE between the DRKT prediction and the exact solution:
\begin{equation}
\mathcal{L}_2 = \frac{1}{N_u}\sum_{i=1}^{N_u} \left| \hat{u}_{\text{DRKT}}^{(i)}(\hat{u}, \hat{u}_x, \hat{u}_{xxx}; {\alpha}) - u_E^{(i)} \right|,
\end{equation}
where $\hat{u}_{\text{DRKT}}$ is the simulated solution obtained after the temporal recursion of DRKT, and $\alpha$ denotes the learnable integration coefficients. This stage enables DRKT to establish a mapping relationship between states at adjacent time instants, laying the foundation for subsequent extrapolation inference.

\textbf{Stage 3: Online Fine-Tuning and Extrapolation Evolution.} This stage constitutes the core innovation of the present work. First, the parameters of the DRKT network are frozen, rendering it a stable evolution operator. Starting from the initial time, the well-trained DRKT from Stage~2 is used to iteratively generate the solution function $\hat{u}_{\text{DRKT}}^k$ for the first time step in the extrapolation phase. This result is then employed as the training target to reactivate the DPINN from Stage~1 for further training. Once training is completed, automatic differentiation is utilized to compute $\partial \hat{u}_{\text{DPINN}} / \partial x$ and $\partial^3 \hat{u}_{\text{DPINN}} / \partial x^3$ at the current time step, thereby preparing the input data required for DRKT to generate the predicted value at the next time step. This iterative process is repeated cyclically until the evolutionary inference over the entire time interval is completed.

The key to this stage lies in the introduction of a training mechanism that combines online fine-tuning with a weighted target. At each extrapolation time step, a small number of rapid fine-tuning iterations are performed on the DPINN, making its output approximate the weighted combination of the exact solution and the DRKT prediction:
\begin{equation}
\mathcal{L}_3 = \frac{1}{N_u}\sum_{i=1}^{N_u} \left| \hat{u}_{\text{DPINN}}^{(i)} - \left( \omega \cdot u_E^{(i)} + (1 - \omega) \cdot \hat{u}_{\text{DRKT}}^{(i)}(\hat{u}, \hat{u}_x, \hat{u}_{xxx}; \boldsymbol{\alpha}_{\text{fixed}}) \right) \right|,
\end{equation}
where the weight $\omega$ decays linearly as extrapolation proceeds. The final prediction is generated by a weighted fusion of the fine-tuned DPINN output, the DRKT prediction, and the exact solution, further enhancing the robustness of the model. This design allows the model to fully leverage exact solution information for guidance in the early extrapolation stage, while smoothly transitioning to trust the physically consistent evolution provided by DRKT as the exact solution gradually recedes, thereby effectively suppressing error accumulation during long-term extrapolation.

Through the above three-stage progressive training, DPINN and DRKT form an efficient complementarity and synergy: DPINN is responsible for high-fidelity reconstruction of the physical state at key time instants, while DRKT ensures stable and reliable evolutionary advancement along the temporal direction. Experimental results demonstrate that, when DSWs problems excited by square-root initial condition, this strategy effectively suppresses error accumulation during long-term extrapolation while maintaining favorable numerical stability.

\section{Exact Solution of the KdV Equation under Square-Root Initial Condition for DSWs}
\hspace{1.5em}A step-like initial profile is an idealization of a pulse with a very sharp leading front. We now consider an alternative scenario where, at the wave-breaking moment $t = 0$, the initial disturbance takes a parabolic form \eqref{eq:square-root form}.

Substituting the initial condition into the Hopf equation yields its implicit solution:
\begin{equation}
x - 6ut = -u^{2},
\label{eq:hopf}
\end{equation}
this solution exhibits a multivalued region $0 < x < 9t^{2}$ for $t > 0$. According to dispersive shock wave theory, this multivalued region is to be replaced by an oscillatory structure generated by dispersive effects. At the trailing edge, the dispersive shock wave solution must match the Hopf equation solution, satisfying the condition:
\begin{equation}
w_{3}|_{r_{1} = r_{2}} = -u^{2},\quad u = r_{3}^{-},
\label{eq:trailing_condition}
\end{equation}
where $r_{1}$, $r_{2}$, $r_{3}$ are the Riemann invariants satisfying $r_{1} \leq r_{2} \leq r_{3}$ and $w_{i}$ are velocity functions to be determined. In the limit $m = (r_{2} - r_{1})/(r_{3} - r_{1}) \to 0$, the velocities $w_{i}$ should behave as quadratic functions of the Riemann invariants. The higher-order symmetries of the KdV equation provide such velocity functions, which exhibit power-law dependence on $r_{i}$ as $m \rightarrow 0$. Accordingly, we construct $w_{i}(r)$ as follows:
\begin{equation}
\begin{array}{c}
w_{i} = C\left(1 - \left(\frac{L}{\partial_{i}L}\right)\partial_{i}\right)W_{2},\quad W_{2} = 2s_{2} - \frac{3}{2}s_{1}^{2},\\
s_{2} = r_{1}r_{2} + r_{2}r_{3} + r_{3}r_{1},\quad s_{1} = r_{1} + r_{2} + r_{3},
\end{array}
\label{eq:wi}
\end{equation}
where $C$ is a constant to be determined, $L$ is the wavelength. At the leading edge $x^{+}$, the vanishing mean amplitude requires $r_{1} = 0$ and $r_{2} = r_{3}$. Thus, in this solution, only one Riemann invariant $r_{1}$ remains constant and equal to zero, while the other two invariants vary with $x$ and $t$ and coincide with each other at the leading edge:
\begin{equation}
r_{2} = r_{3},\quad \text{that is},\quad m = \frac{r_{2}}{r_{3}} = 1\quad \text{at}\quad x = x^{+}.
\label{eq:leading_edge}
\end{equation}
At the trailing edge, oscillations disappear, corresponding to the condition:
\begin{equation}
r_{2} = r_{1} = 0,\quad \text{that is},\quad m = \frac{r_{2}}{r_{3}} = 0\quad \text{at}\quad x = x^{-}.
\label{eq:trailing_edge}
\end{equation}
The Whitham equation for $r_{1}$ is satisfied identically, whereas $r_{2}$ and $r_{3}$ are determined by:
\begin{equation}
x - v_{2}t = w_{2},\quad x - v_{3}t = w_{3},
\label{eq:whitham}
\end{equation}
taking $r_{1} = 0$, the expressions for $v_{2}$ and $v_{3}$ simplify to:
\begin{equation}
\begin{array}{c}
v_{2} = 2r_{3}\left(1 + m - \frac{2m(1 - m)K(m)}{E(m) - (1 - m)K(m)}\right),\\
v_{3} = 2r_{3}\left(1 + m + \frac{2(1 - m)K(m)}{E(m)}\right),
\end{array}
\label{eq:v_velocities}
\end{equation}
and $w_{2}$ and $w_{3}$ are given by Eq.~\eqref{eq:wi} with $r_{1} = 0$:
\begin{equation}
\begin{array}{l}
w_{2} = C r_{3}^{2}\left(2m - \frac{3}{2}(1 + m)^{2} + \frac{2m(1 - m^{2})K(m)}{E(m) - (1 - m)K(m)}\right),\\
w_{3} = C r_{3}^{2}\left(2m - \frac{3}{2}(1 + m)^{2} - \frac{2(1 - m)(3 + m)K(m)}{E(m)}\right),
\end{array}
\label{eq:w_velocities}
\end{equation}
the constant $C$ is determined by the boundary condition. Since $w_{3} = -\frac{15}{2}C r_{3}^{2}$ as $m \rightarrow 0$, we obtain $C = 2/15$. Substituting this value yields the final expressions for the solution of the Whitham equations:
\begin{equation}
\begin{array}{l}
x - v_{2}t = \frac{2}{15}\left[W + \left(\frac{1}{2}v_{2} - s_{1}\right)\partial W / \partial r_{2}\right],\\
x - v_{3}t = \frac{2}{15}\left[W + \left(\frac{1}{2}v_{3} - s_{1}\right)\partial W / \partial r_{3}\right],
\end{array}
\label{eq:whitham_solution}
\end{equation}
where $W = 2r_{2}r_{3} - \frac{3}{2}(r_{2} + r_{3})^{2}$ and $s_{1} = r_{2} + r_{3}$.

Near the trailing edge, as $m \rightarrow 0$, the above two equations reduce to:
\begin{equation}
x^{-} + 6r_{3}^{-}t = \frac{1}{3}(r_{3}^{-})^{2},\quad x^{-} - 6r_{3}^{-}t = -(r_{3}^{-})^{2},
\label{eq:trailing_reduce}
\end{equation}
this gives the parametric representation of the trailing edge motion:
\begin{equation}
x^{-} = -\frac{1}{3}(r_{3}^{-})^{2},\quad t = \frac{1}{9}r_{3}^{-},
\label{eq:trailing_param}
\end{equation}
eliminating the parameter $r_{3}^{-}$ yields:
\begin{equation}
x^{-} = -27t^{2}.
\label{eq:trailing_law}
\end{equation}

At the leading edge, as $m \rightarrow 1$ and $r_{2} = r_{3}$, equation tends to:
\begin{equation}
x^{+} - 4r_{3}t = -\frac{8}{15}r_{3}^{2},
\label{eq:leading_reduce}
\end{equation}
the value of $r^{+}$ is determined by the condition that $x$ attains its maximum within the oscillation region, i.e., $dx^{+}/dr_{3}|_{t = const} = 0$, which yields $r_{3} = 15t/4$. Substituting this gives:
\begin{equation}
x^{+} = \frac{15}{2}t^{2}.
\label{eq:leading_law}
\end{equation}
This is the law of motion for the leading edge.

\begin{figure}[t!]
\centering
\includegraphics[scale=0.4]{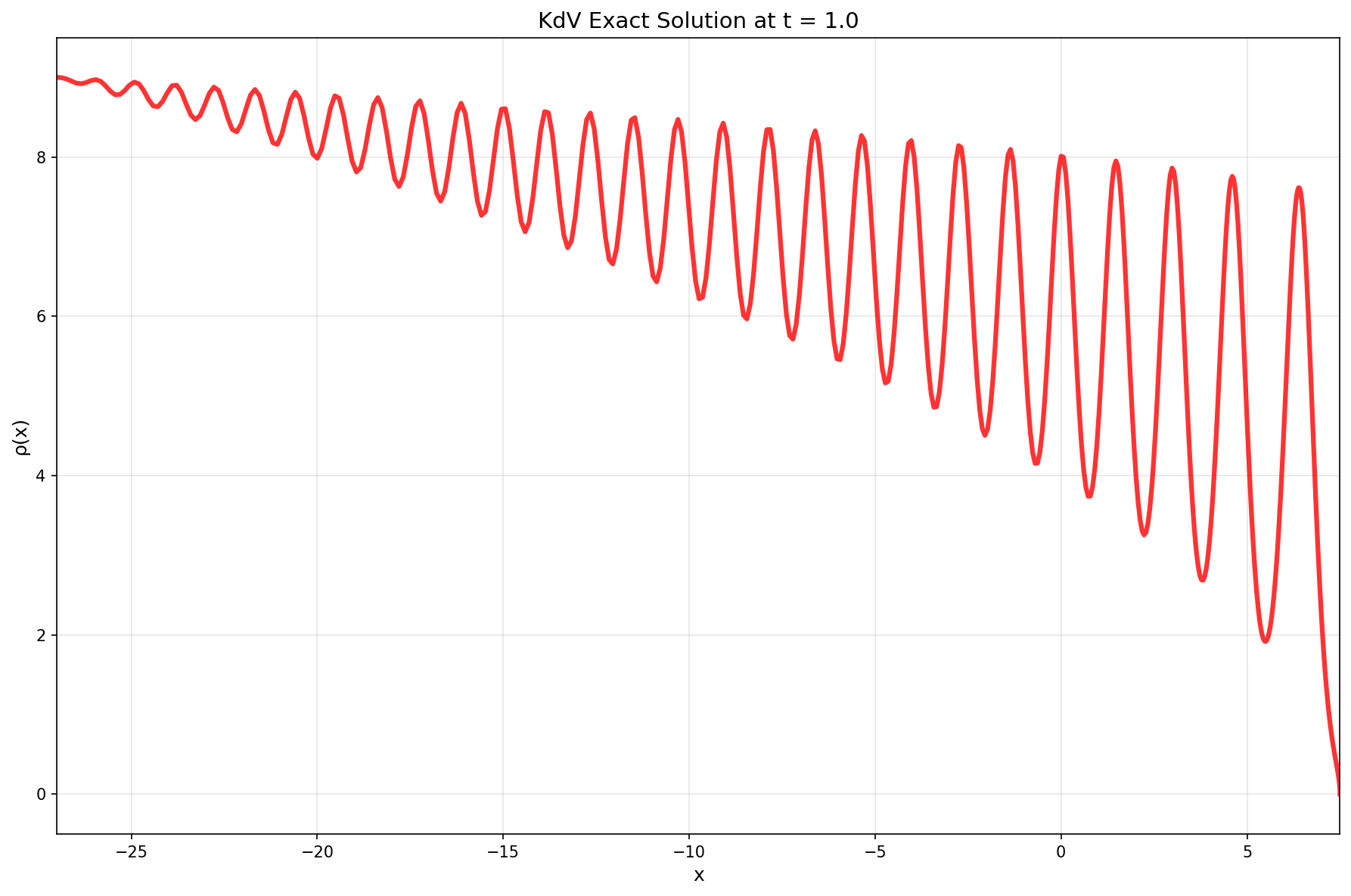}

\vspace{-0.2cm}
{\footnotesize \textbf{Fig.~5.} DSWs of the KdV equation under square-root initial condition at $t = 1$.}
\end{figure}

From Eqs.~\eqref{eq:trailing_law} and \eqref{eq:leading_law}, the propagation velocities of the edges are obtained as:
\begin{equation}
v_{trailing} = \frac{dx^{-}}{dt} = -54t,\quad v_{leading} = \frac{dx^{+}}{dt} = 15t.
\label{eq:velocities}
\end{equation}
For the generalized KdV equation:
\begin{equation}
u_{t} + \alpha u u_{x} + u_{xxx} = 0,
\label{eq:generalized_kdv}
\end{equation}
the nonlinear coefficient $\alpha = 6$ in the above derivation appears in the velocity expressions. Through dimensional analysis, the relationship between the edge velocities and the equation parameter is obtained as:
\begin{equation}
v_{trailing} = \alpha \cdot r_{3}^{-} = -9 \alpha t,
\label{eq:trailing_alpha}
\end{equation}
\begin{equation}
v_{leading} = \frac{\alpha}{3}(r_{1}^{+} + 2r_{3}^{+}),
\label{eq:leading_alpha}
\end{equation}
at $t = 1$, with $r_{1}^{+} = 0$ and $r_{3}^{+} = \frac{15}{4}t$, we have:
\begin{equation}
v_{leading} = \frac{5\alpha t}{2}.
\label{eq:leading_alpha_t1}
\end{equation}
When $\alpha = 6$, this gives $v_{leading} = 15$, consistent with the numerical result. Example of such a plot $u(x,t)$ at fixed value of $t$ is shown in Fig.~5.

\section{Experimental Results of Different Models}

\subsection{Std-PINN for the KdV Equation}
\hspace{1.5em}This section employs the Std-PINN to solve the evolution problem of the KdV equation under square-root initial condition, and compares its predictions with the exact solution to evaluate the practical performance of this method in handling such dispersive nonlinear wave problems.

Fig.~6 presents the prediction results of the Std-PINN over the entire spatiotemporal domain and their comparison with the exact solution. Fig.~6~(a) displays the Std-PINN predicted solution, the exact reference solution, and the absolute error heatmap between them. From the overall spatiotemporal evolution plots, it can be clearly observed that, even within the training region (the first 900 time steps), the Std-PINN fails to accurately capture the complex structure of the DSWs. The predicted results exhibit a pronounced over-smoothing characteristic, with the high-frequency oscillatory components near the wavefront being almost entirely lost, preserving only the rough envelope of the wave packet. As the evolution progresses, the discrepancy between the model predictions and the exact solution continues to widen. Entering the later stage of the training region, the predictions deviate significantly from the true solution, with the fine oscillatory structures at the wavefront virtually vanishing. In the extrapolation region (the last 100 time steps), the model performance deteriorates further, yielding errors of orders of magnitude between the outputs and the ground truth. 

\begin{figure}[t!]
\centering

\includegraphics[scale=0.35]{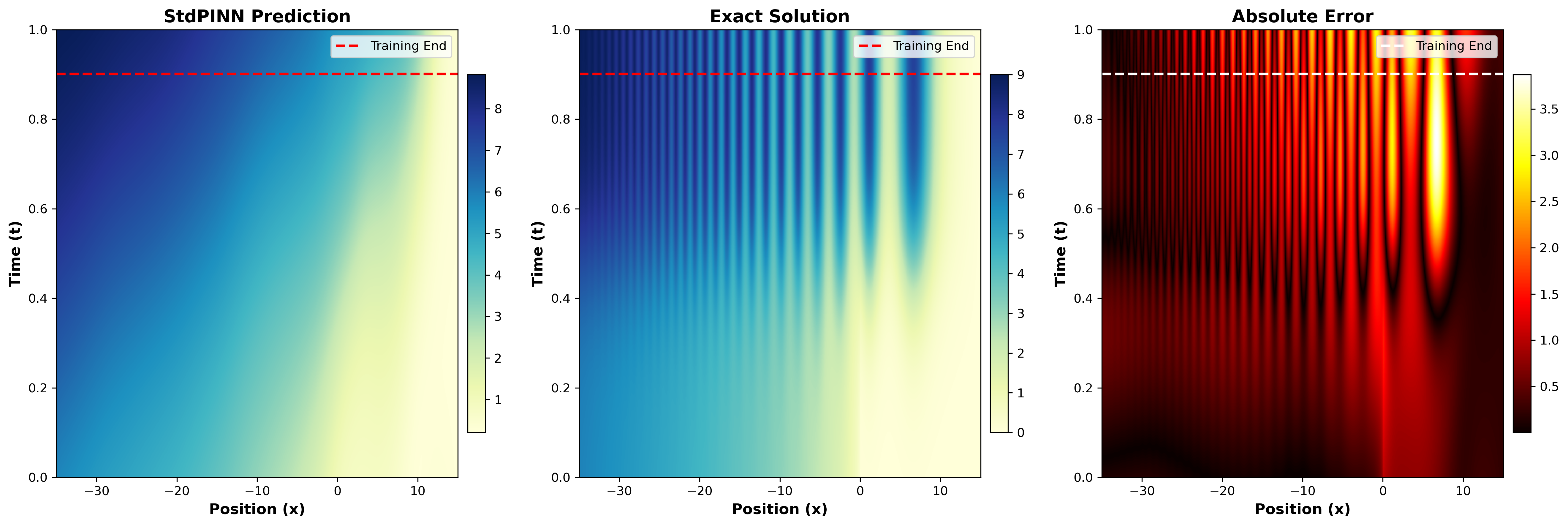} \\
\vspace{0.1cm}
{\footnotesize (a)}

\vspace{0.5cm}

\includegraphics[scale=0.2]{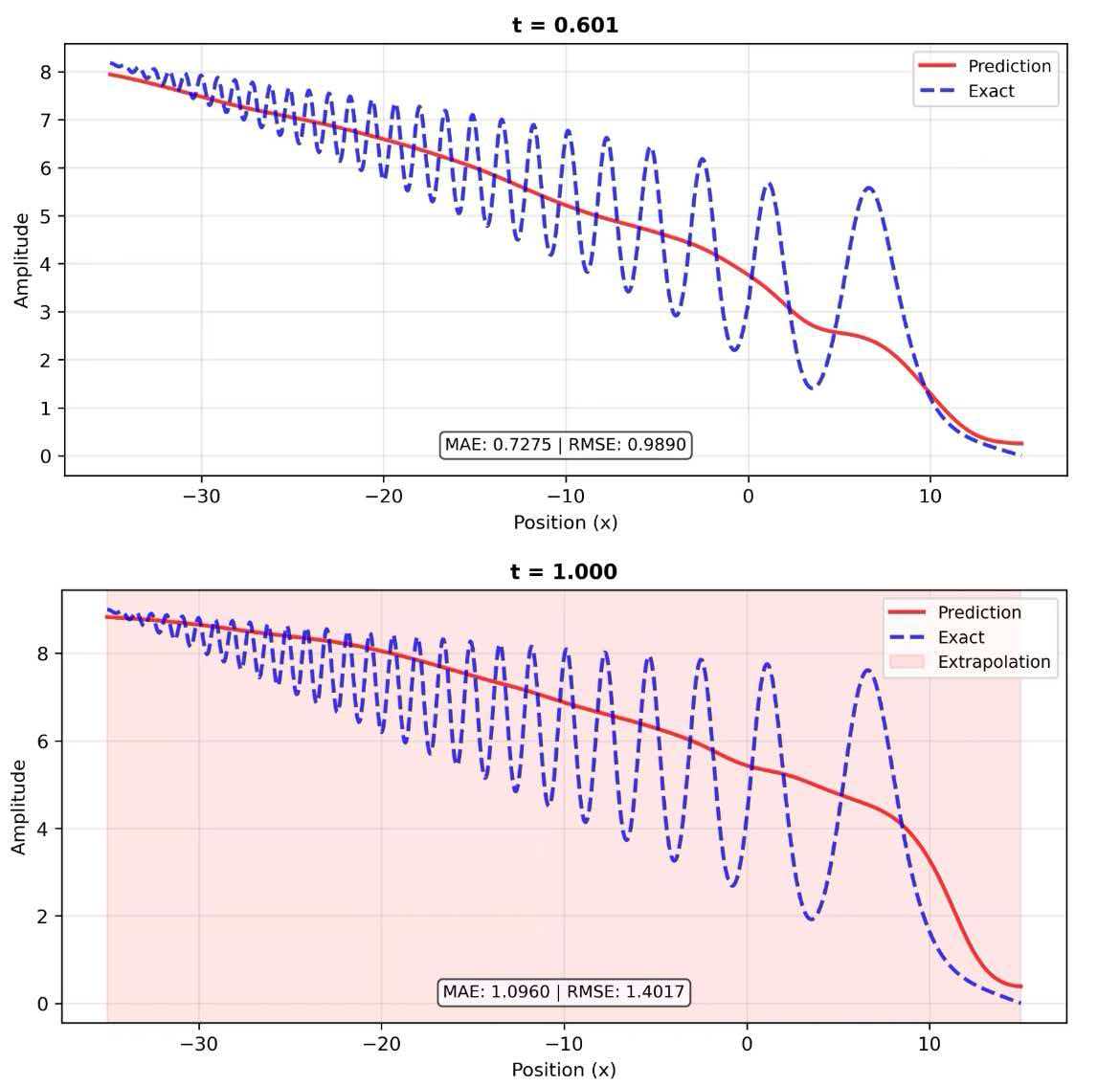} \\
\vspace{0.1cm}
{\footnotesize (b)}

\vspace{0.3cm}
{\footnotesize
\textbf{Fig.~6.} (a) Std-PINN solution of the KdV equation over the entire spatiotemporal domain, reference solution, and their absolute error heatmap. (b) Time snapshots of the Std-PINN solution and the reference solution at $t=0.601$ (Top) and $t=1.000$ (Bottom).
}
\end{figure}

Fig.~6~(b) further provides a comparison of time snapshots at two representative instants ($t = 0.601$ and $t = 1.000$). At $t = 0.601$, the Std-PINN prediction can still roughly follow the overall trend of the wave packet, but the oscillatory details at the wavefront have already undergone noticeable attenuation. At $t = 1.000$, the discrepancy between the predicted result and the exact solution has become substantially pronounced, and the model has almost entirely lost the ability to describe the wave behavior in the extrapolation phase.

The above results reveal several fundamental limitations of the Std-PINN when solving nonlinear evolution equations with dispersive effects. First, the physical residual loss only constrains local differential relations, making it difficult to maintain the phase consistency of oscillatory structures over long time scales. Second, the inherent spectral bias of neural networks causes them to preferentially learn low-frequency components, with severely insufficient capacity for learning high-frequency oscillatory components. Third, the competition among different loss terms within the multi-task learning framework hinders the model from achieving an effective balance among initial condition fitting, boundary constraint enforcement, and equation residual minimization. Taking all these observations together, we conclude that the Std-PINN is inadequate for providing sufficiently accurate numerical simulations when confronted with complex nonlinear wave systems such as DSWs that are characterized by rich oscillatory features.

\subsection{PgMSNN for the KdV Equation}
\hspace{1.5em}This section employs the PgMSNN to solve the evolution problem of the KdV equation under square-root initial condition. This PgMSNN framework was successfully applied to the step initial value problem of the generalized Gardner equation and serves as a baseline model to evaluate the necessity of the improvements introduced in this paper. It should be noted that PgMSNN and IPgMSNN, which will be introduced later, share the same structure in the first two training stages. The core difference lies in the third stage: PgMSNN lacks the online fine-tuning and weighted target mechanisms, whereas IPgMSNN incorporates these strategies to suppress error accumulation during long-term extrapolation.

\begin{figure}[t!]
\centering

\includegraphics[scale=0.35]{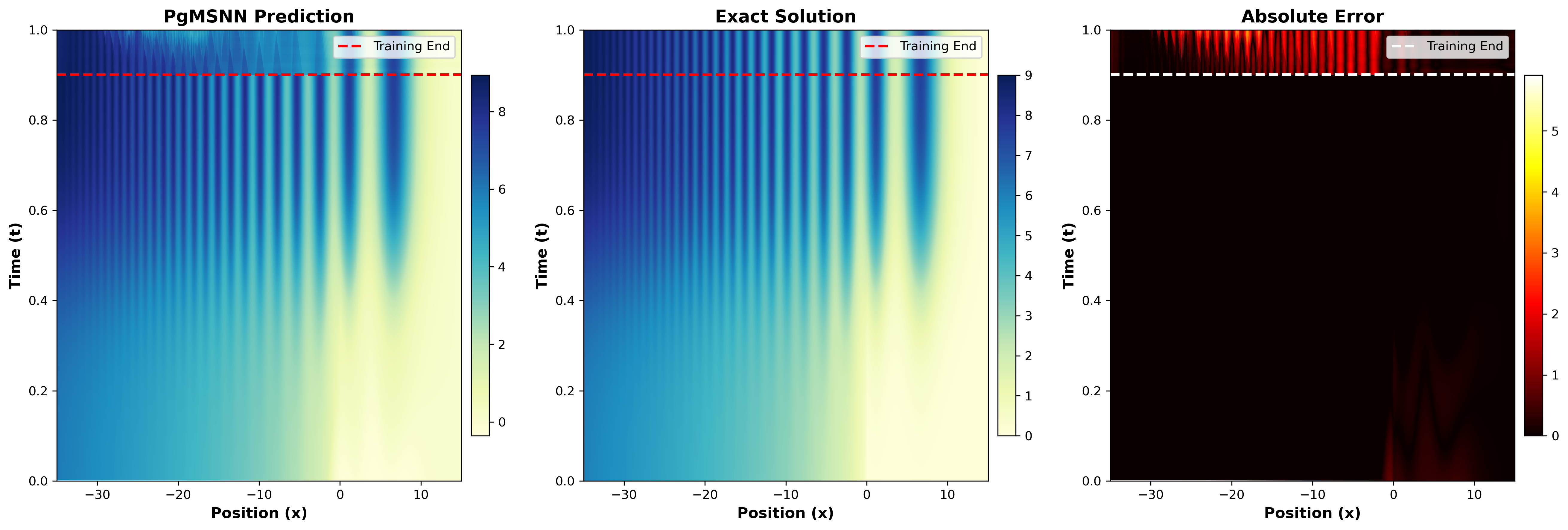} \\
\vspace{0.1cm}
{\footnotesize (a)}

\vspace{0.5cm}

\includegraphics[scale=0.2]{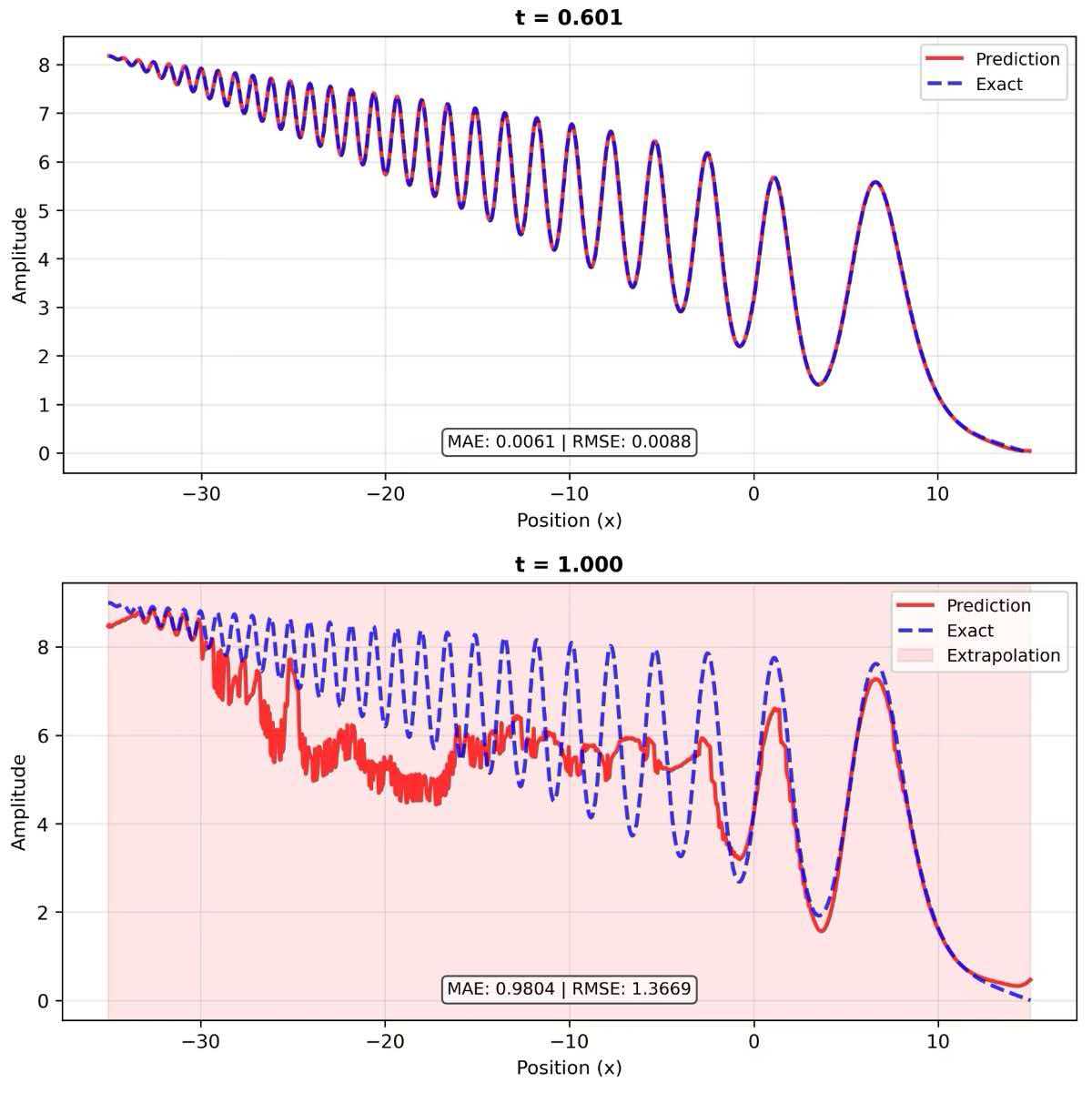} \\
\vspace{0.1cm}
{\footnotesize (b)}

\vspace{0.3cm}
{\footnotesize
\textbf{Fig.~7.} (a) PgMSNN solution of the KdV equation over the entire spatiotemporal domain, reference solution, and their absolute error heatmap. (b) Time snapshots of the PgMSNN solution and the reference solution at $t=0.601$ (Top) and $t=1.000$ (Bottom).
}
\end{figure}

Fig.~7 presents the prediction results of PgMSNN over the entire spatiotemporal domain and their comparison with the exact solution. Fig.~7~(a) displays the PgMSNN predicted solution, the exact reference solution, and the absolute error heatmap between them. From the overall spatiotemporal evolution plots, it can be observed that within the training region, PgMSNN is able to reconstruct the wave packet structure of the DSWs reasonably well. However, upon entering the extrapolation region, the model performance degrades significantly, with the high-frequency oscillatory components near the wavefront gradually being lost and errors accumulating rapidly.

Fig.~7~(b) further provides a comparison of time snapshots at two representative instants \(t = 0.601\) and \(t = 1.000\). At \(t = 0.601\), which lies within the training region, the PgMSNN predictions agree well with the exact solution, with both the amplitude and phase of the high-frequency oscillations accurately reproduced. This validates the effectiveness of the first two stages of PgMSNN. However, at \(t = 1.000\), which lies in the extrapolation region, the discrepancy between the predicted results and the exact solution becomes substantial, and the fine oscillatory structures at the wavefront are largely lost, preserving only a rough envelope.

These results demonstrate that although PgMSNN achieves satisfactory performance within the training region through the introduction of the dispersion factor and the Deep Runge-Kutta scheme, its lack of an online fine-tuning mechanism leads to rapid error accumulation during extrapolation, making it difficult to maintain wavefront details over long-term evolution. This limitation motivates the improvement introduced in the third stage of our framework, namely the incorporation of the weighted target mechanism and online fine-tuning strategy, leading to the IPgMSNN framework, which will be discussed in the following subsection.

\subsection{IPgMSNN for the KdV Equation}
\hspace{1.5em}This section employs the IPgMSNN to solve the evolution problem of the KdV equation under square-root initial condition. The prediction results of IPgMSNN are compared with the exact solution, as well as with the results of the Std-PINN and PgMSNN, to validate the effectiveness and superiority of the improved method proposed in this paper.

Fig.~8 presents the prediction results of IPgMSNN over the entire spatiotemporal domain and their comparison with the exact solution. Fig.~8~(a) displays the IPgMSNN predicted solution, the exact reference solution, and the absolute error heatmap between them. From the overall spatiotemporal evolution plots, it can be clearly observed that IPgMSNN not only accurately reproduces the complex structure of the DSWs within the training region, but also maintains satisfactory predictive capability in the extrapolation region. The high-frequency oscillatory components near the wavefront are effectively captured, and both the envelope morphology of the wave packet and its internal oscillatory details remain in excellent agreement with the exact solution. From the error heatmap, the error distribution is relatively uniform, primarily concentrated in the trailing edge region.

\begin{figure}[t!]
\centering

\includegraphics[scale=0.35]{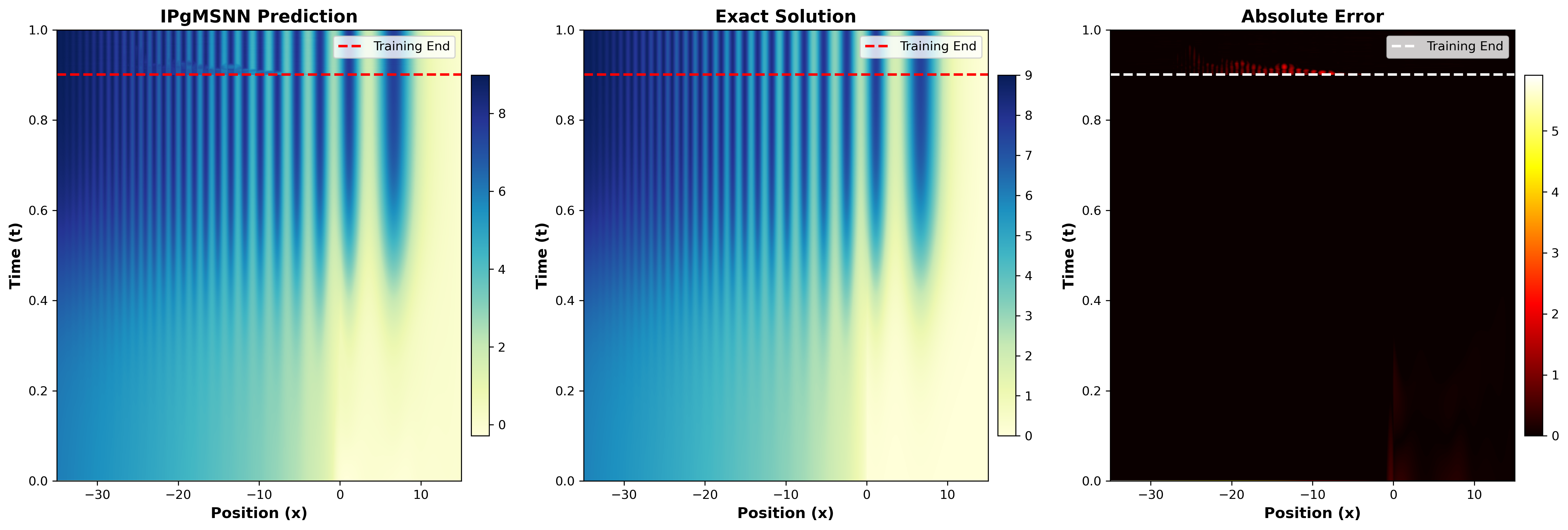} \\
\vspace{0.1cm}
{\footnotesize (a)}

\vspace{0.5cm}

\includegraphics[scale=0.2]{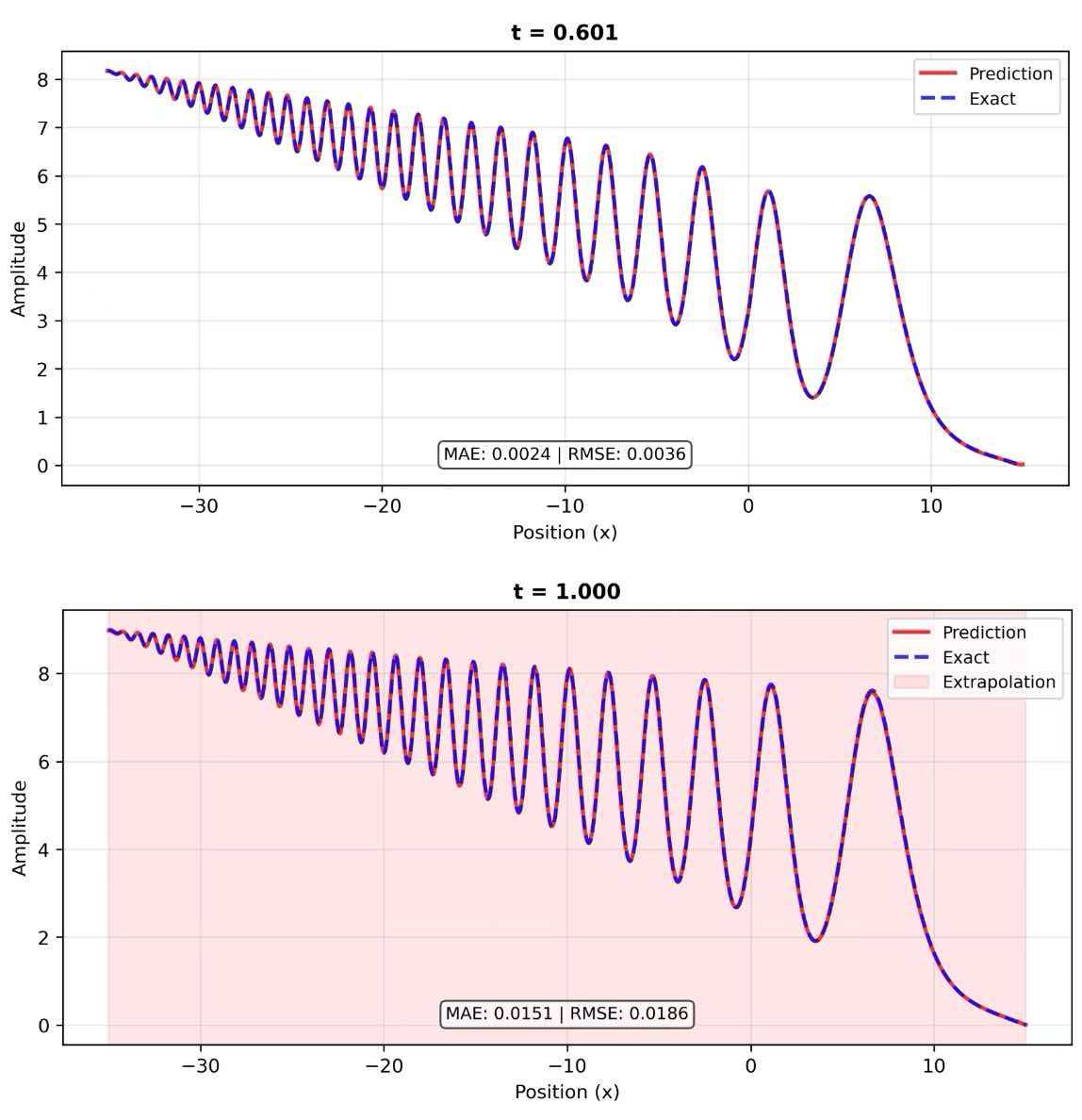} \\
\vspace{0.1cm}
{\footnotesize (b)}

\vspace{0.3cm}
{\footnotesize
\textbf{Fig.~8.} (a) IPgMSNN solution of the KdV equation over the entire spatiotemporal domain, reference solution, and their absolute error heatmap. (b) Time snapshots of the IPgMSNN solution and the reference solution at $t=0.601$ (Top) and $t=1.000$ (Bottom).
}
\end{figure}

Fig.~8~(b) further provides a comparison of time snapshots at two representative instants \(t = 0.601\) and \(t = 1.000\). At \(t = 0.601\), which lies within the training region, the IPgMSNN predictions agree well with the exact solution, with both the amplitude and phase of the wavefront oscillations accurately reproduced. At \(t = 1.000\), which lies in the extrapolation region, the IPgMSNN prediction still maintains favorable consistency with the exact solution. Although the wavefront details exhibit slight attenuation, the overall oscillatory structure remains clearly discernible. In contrast, the Std-PINN predictions exhibit a pronounced over-smoothing characteristic, with the high-frequency oscillatory components being almost entirely lost. Meanwhile, PgMSNN shows significant performance degradation in the extrapolation region, with the wavefront oscillatory details gradually being lost. The growth rate of the prediction error of IPgMSNN as evolution progresses is substantially slower than that of both the Std-PINN and PgMSNN, demonstrating the excellent stability of the model during long-term extrapolation.

The above results fully validate the effectiveness of the three-stage progressive training strategy employed in the IPgMSNN model. Through the first stage of high-precision reconstruction of the initial wave packet, the model accurately captures the key spectral information excited by the square-root initial condition. Through the second stage of training the DRKT network, the model masters the dynamical evolution laws described by the KdV equation. Through the third stage of combining online fine-tuning with a weighted target mechanism, the model achieves a smooth transition from exact solution guidance to physically consistent evolution during extrapolation, effectively suppressing error accumulation in long-term extrapolation. Taken together, these results demonstrate that IPgMSNN, when confronted with complex nonlinear wave systems such as DSWs that are characterized by rich oscillatory features, exhibits simulation accuracy and extrapolation stability that far surpass those of the Std-PINN and PgMSNN.

\section{Model Stability Analysis}
\hspace{1.5em}In practical applications, training data inevitably contain a certain degree of measurement noise or observational errors. Therefore, evaluating the robustness of the model under conditions of compromised data quality is of significant importance for validating its applicability in real-world scenarios. This section systematically investigates the stability performance of the IPgMSNN model in a noisy environment by imposing $1\%$ Gaussian noise on the training data. The experimental configuration remains consistent with that in Section~5, with 1000 sampling points in both the spatial and temporal domains, the first 900 time steps used for training, and the remaining 100 time steps reserved for extrapolation testing.

\begin{figure}[t!]
\centering

\includegraphics[scale=0.35]{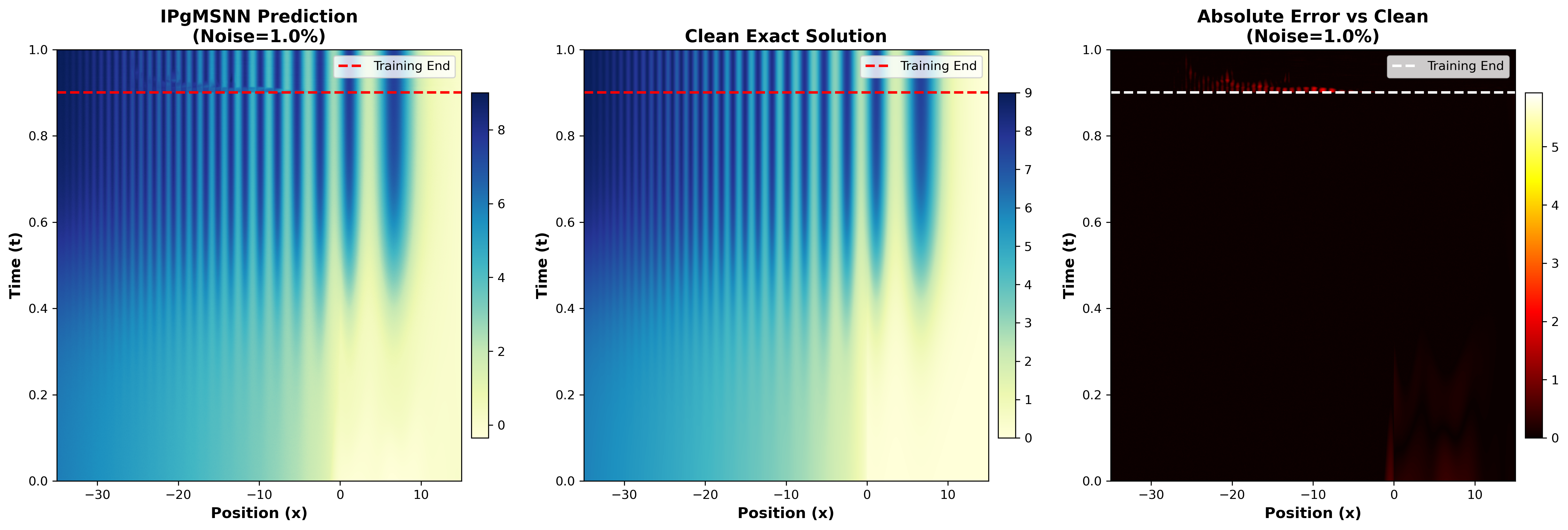} \\
\vspace{0.1cm}
{\footnotesize (a)}

\vspace{0.5cm}

\includegraphics[scale=0.2]{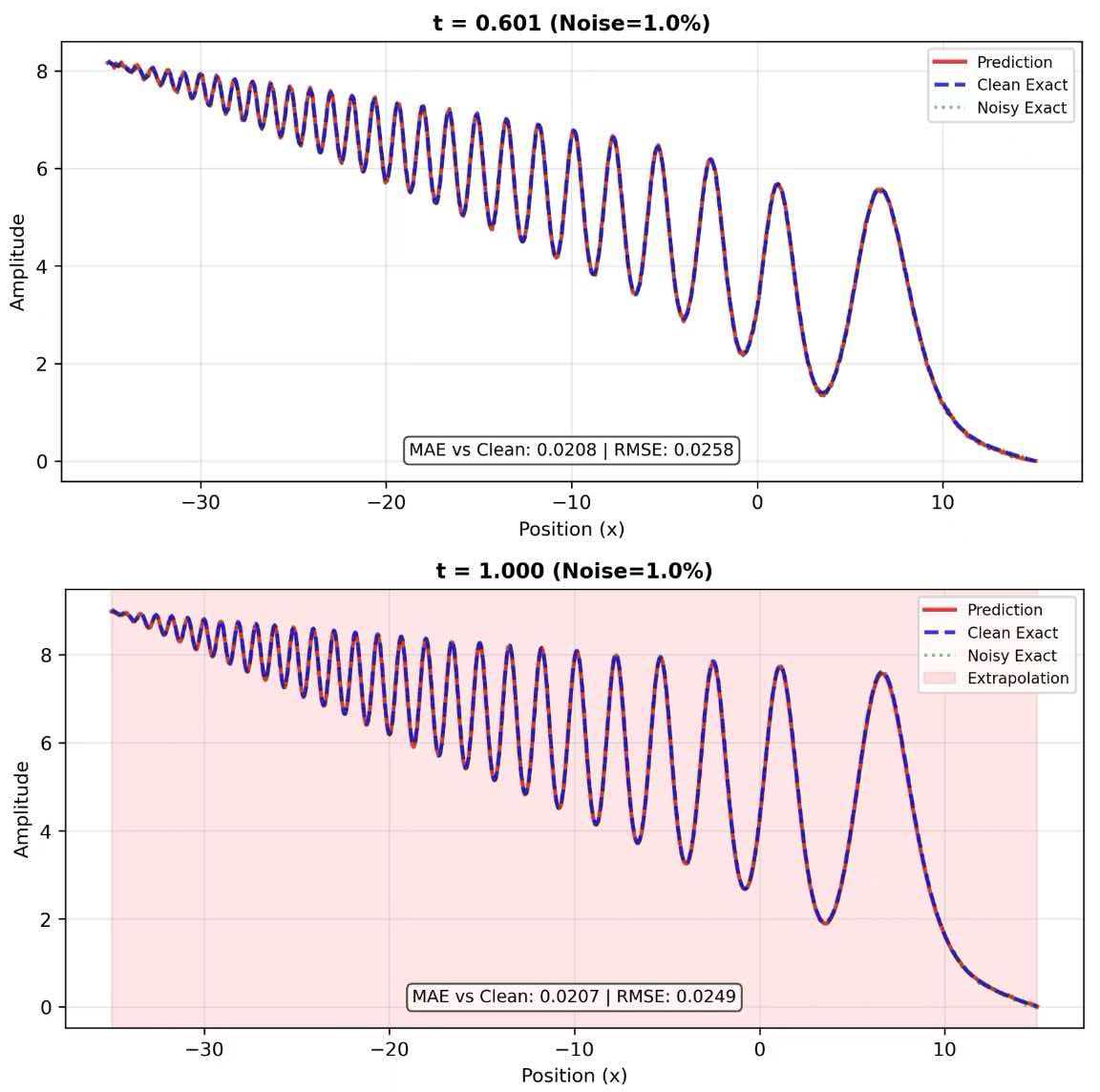} \\
\vspace{0.1cm}
{\footnotesize (b)}

\vspace{0.3cm}
{\footnotesize
\textbf{Fig.~9.} (a) IPgMSNN solution of the KdV equation over the entire spatiotemporal domain trained with 1\% Gaussian noise, reference solution, and their absolute error heatmap. (b) Time snapshots of the IPgMSNN solution and the reference solution at $t=0.601$ (Top) and $t=1.000$ (Bottom).
}
\end{figure}

Fig.~9 presents the prediction results of IPgMSNN under noisy training data. Fig.~9~(a) displays the IPgMSNN predicted solution, the noise-free exact reference solution, and the absolute error heatmap between them. From the overall spatiotemporal evolution plots, it can be observed that, despite the introduction of $1\%$ Gaussian noise into the training data, IPgMSNN is still capable of reasonably reproducing the overall structure of the DSWs. The main morphology of the wave packet and the essential characteristics of the wavefront oscillations are effectively preserved, and the prediction results do not exhibit divergence or severe distortion due to noise interference. From the error heatmap, the error distribution is relatively uniform across the entire spatiotemporal domain, with no localized sharp amplification of errors induced by the introduction of noise.

Fig.~9~(b) further provides a comparison of time snapshots at two representative instants ($t = 0.601$ and $t = 1.000$). At $t = 0.601$ within the training region, despite the presence of noise in the training data, the IPgMSNN predictions maintain favorable agreement with the noise-free exact solution, with both the phase and amplitude of the wavefront oscillations being accurately reproduced. At $t = 1.000$ in the extrapolation region, the oscillatory structure of the wavefront is preserved to a certain extent, and the rate of error accumulation is relatively slow, indicating that the generalization capability of the model in unsupervised regions does not suffer significant degradation due to noise contamination of the training data.

From a quantitative perspective, on the training set, the MAE between the model predictions and the noise-free exact solution remains on the same order of magnitude as the training error under noise-free conditions. On the extrapolation set, the error similarly remains on the order of $10^{-2}$, essentially comparable to the extrapolation error in the absence of noise. Overall, the MAE of the model over the entire spatiotemporal domain is on the order of $10^{-2}$, with an error increase of no more than $30\%$ compared to the noise-free case.

The above results demonstrate that the IPgMSNN model possesses strong tolerance to a certain level of noise interference. Even when the data quality is compromised, the model is still capable of providing reliable numerical simulations. This robustness can be attributed to the synergistic operation of multiple components within the three-stage training strategy: the high-fidelity reconstruction of the initial wave packet structure by DPINN provides a stable starting point for subsequent evolution; DRKT learns the deterministic evolution law described by the KdV equation rather than fitting the noise in the training data, thus exhibiting a certain robustness to random perturbations; the online fine-tuning mechanism in the third stage continuously corrects deviations introduced by noise during the evolution process.

\section{Inverse Problem}
\hspace{1.5em}In the numerical simulation of PDEs, the core objective of the inverse problem lies in estimating or inferring unknown parameters within the equations based on limited observational data. This section systematically investigates the parameter inversion performance of the IPgMSNN model for the evolution problem of the KdV equation under square-root initial condition, and compares it with the PgMSNN and Std-PINN models. The KdV equation under consideration adopts the generalized form $u_t + \alpha u u_x + u_{xxx} = 0$, with the target parameter for inversion being the nonlinear coefficient $\alpha$, whose true value is $6$.

\begin{figure}[t!]
\centering
\includegraphics[scale=0.3]{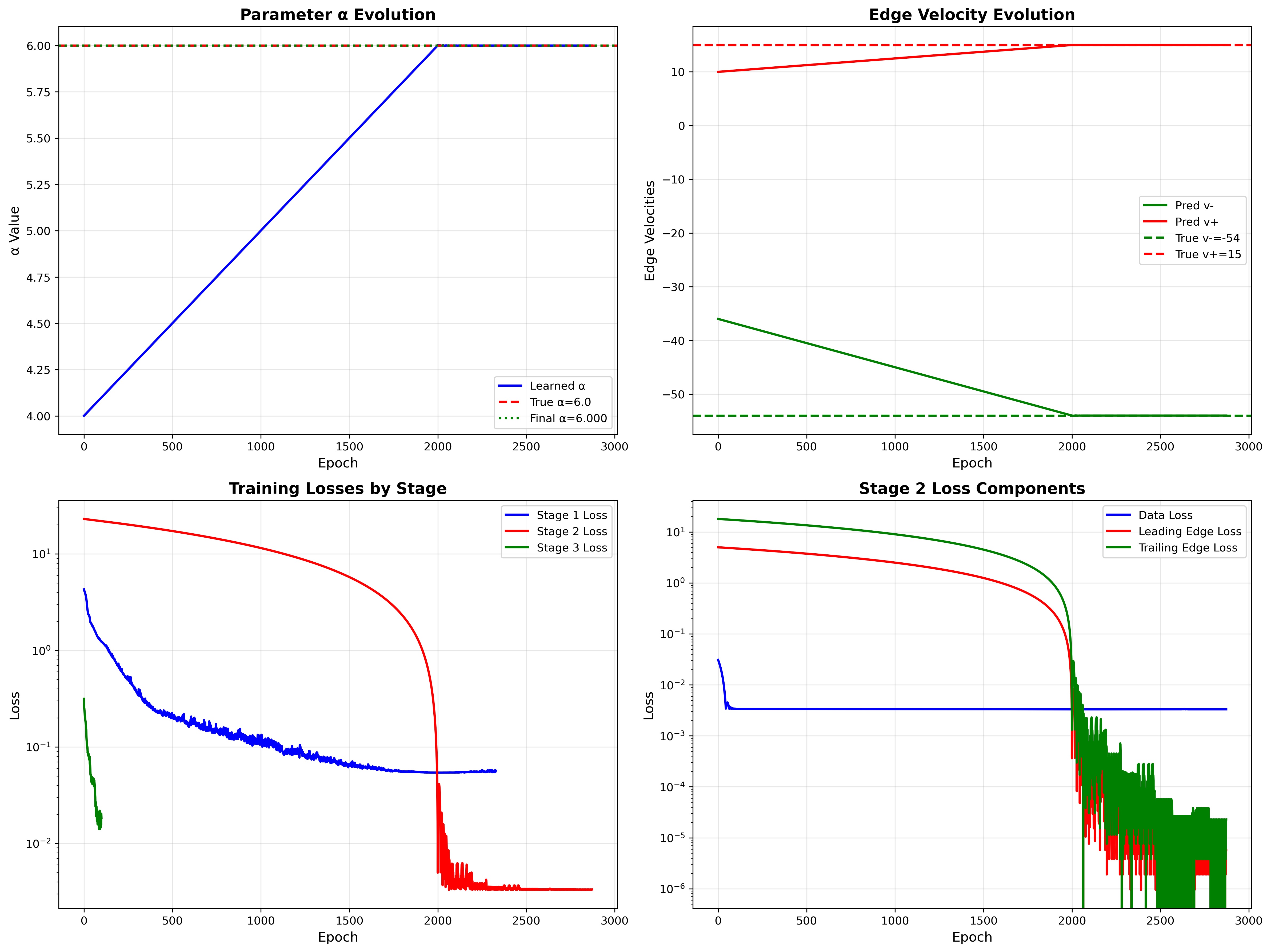} \\
{\footnotesize (a)}

\vspace{0.5cm}

\includegraphics[scale=0.3]{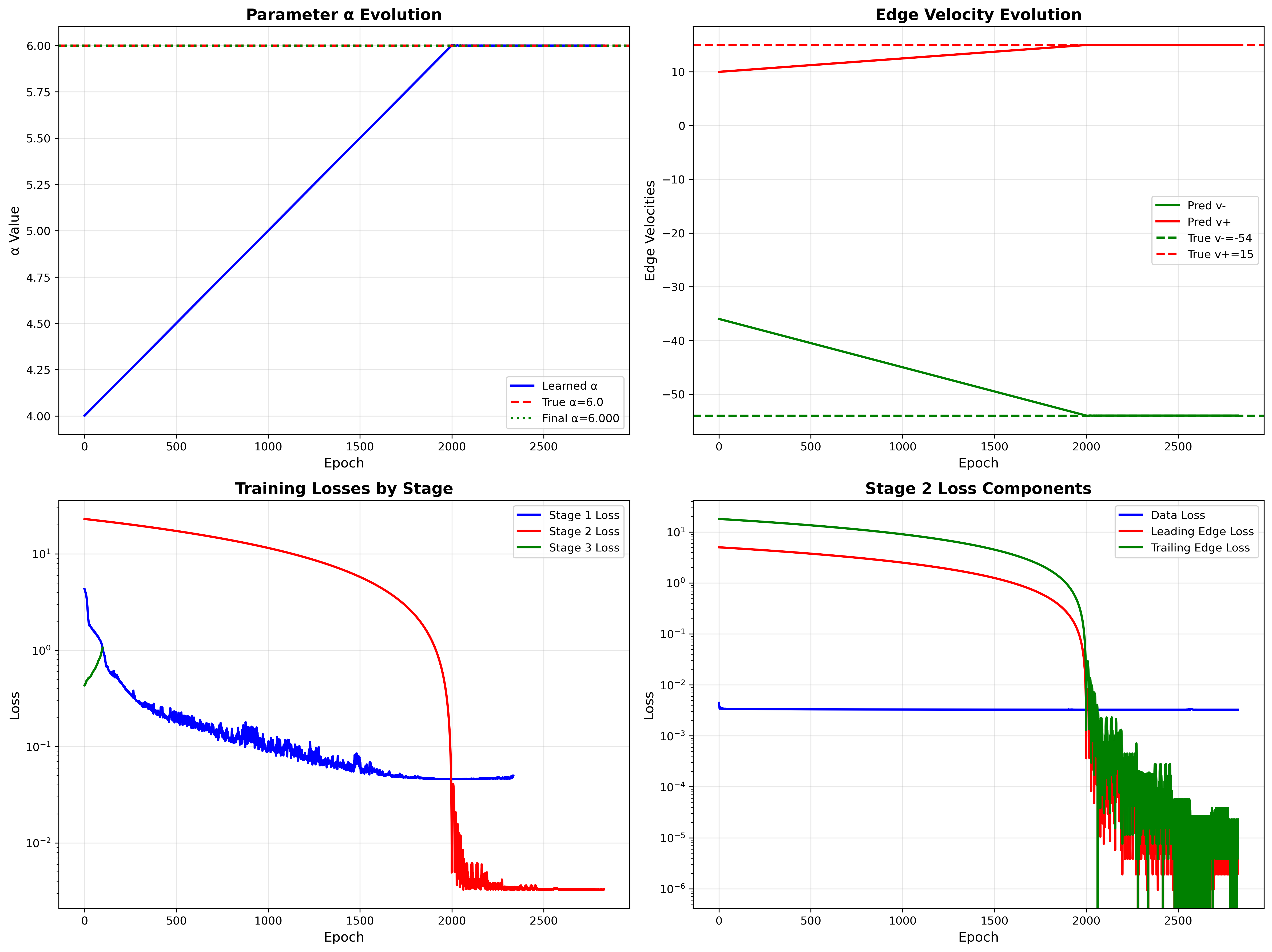} \\
{\footnotesize (b)}

\vspace{0.3cm}
{\footnotesize
\textbf{Fig.~10.} Comparison of parameter inversion and loss curves of IPgMSNN (a), PgMSNN (b), and Std-PINN (c) in the inverse problem.
}
\end{figure}

\begin{figure}[t!]
\centering
\includegraphics[scale=0.3]{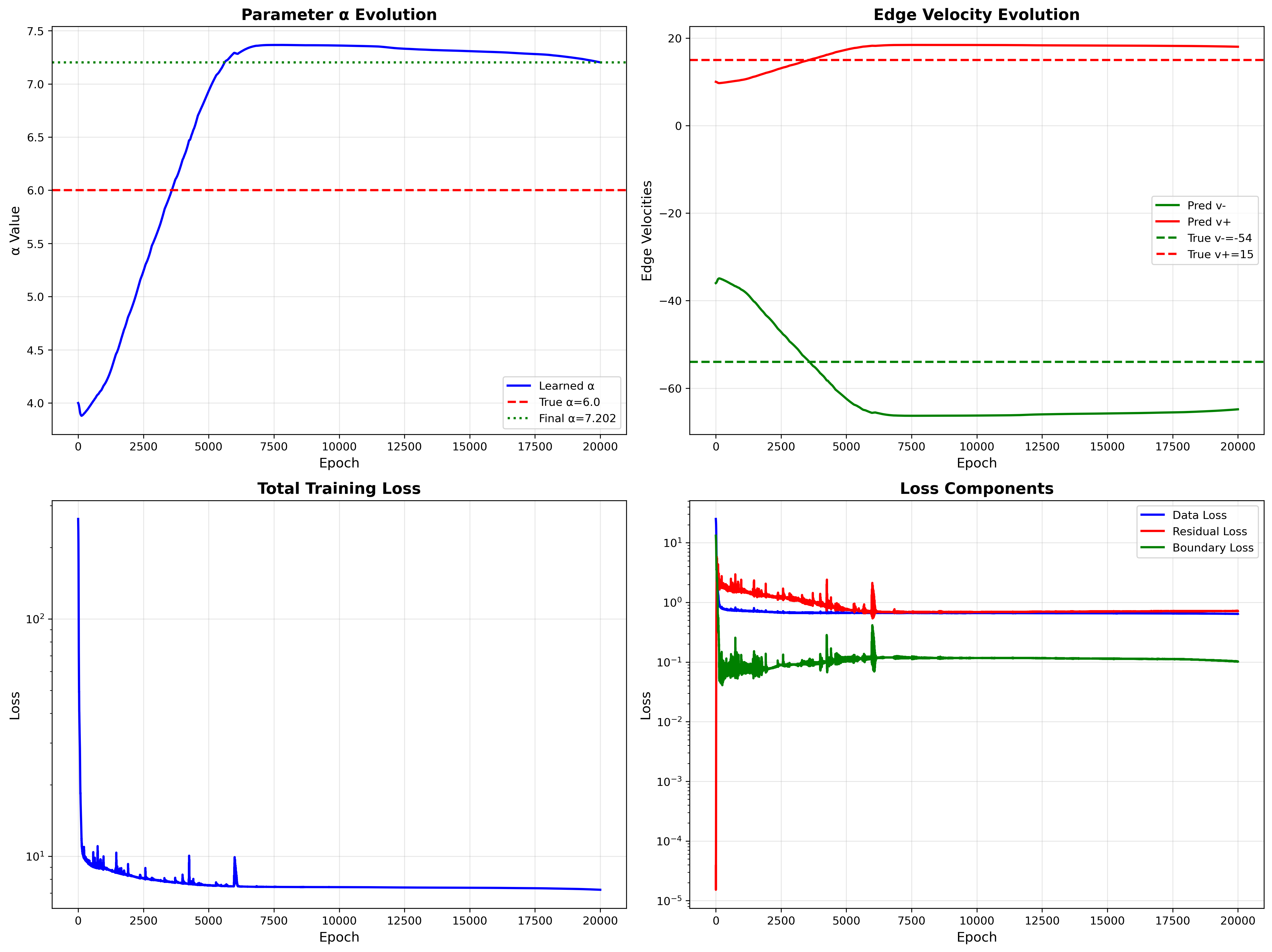} \\
{\footnotesize (c)}

\vspace{0.3cm}
{\footnotesize
\textbf{Fig.~10.} (continued)
}
\end{figure}

Within the IPgMSNN and PgMSNN frameworks, the equation coefficient to be inverted is transformed into a trainable parameter, with the learning task assigned to the DRKT network in the second stage. During this stage, the DRKT network not only learns the temporal evolution dynamics but also optimizes the coefficient $\alpha$, ensuring consistency with both data fitting and the physical constraints imposed by the leading and trailing edge characteristic velocities. For the Std-PINN, the coefficient $\alpha$ is also treated as a trainable parameter optimized alongside the network weights, with the loss function comprising both data fitting and physical residual terms. The experimental configuration remains consistent with Section~5, with 1000 sampling points in both the spatial and temporal domains, the first 900 time steps used for training, and the remaining 100 time steps reserved for extrapolation testing.

The loss function for the inverse problem in the second stage consists of two components: the data fitting loss and the characteristic velocity constraint loss:
\begin{equation}
Loss_2 = Loss_{data} + Loss_{CV},
\end{equation}
where $Loss_{data}$ represents the data fitting loss between the DRKT predictions and the exact solution, employing the MAE:
\begin{equation}
Loss_{data} = \frac{1}{N_u} \sum_{i=1}^{N_u} \left| \hat{u}_{DRKT}^i(\hat{u}, \hat{u}_x, \hat{u}_{xxx}; \alpha) - u_E^i \right|,
\end{equation}
and $Loss_{CV}$ denotes the constraint loss for the leading and trailing edge characteristic velocities, comprising two terms:
\begin{equation}
Loss_{CV} = |v_{leading}(\alpha) - v_{leading}^{true}| + |v_{trailing}(\alpha) - v_{trailing}^{true}|.
\end{equation}
The theoretical relationship between the edge velocities and the equation parameter $\alpha$ has been derived in Section~4. By incorporating these physical constraints into the loss function, the DRKT network is able to learn the equation coefficient more accurately under limited data conditions while maintaining satisfactory extrapolation capability.

Fig.~10 compares the parameter inversion and loss curves of the IPgMSNN, PgMSNN, and Std-PINN models in the inverse problem of the KdV equation under square-root initial condition. Each subfigure consists of four parts: parameter $\alpha$ evolution, edge velocity evolution, training losses by stage, and loss components. Fig.~10~(a) presents the inverse problem results of the IPgMSNN model. From the evolution curve of parameter $\alpha$, it can be observed that $\alpha$ starts from the initial value of 4.0 and rapidly converges to the true value of 6.0 as the number of training epochs increases, remaining stable thereafter. The edge velocity evolution curves show that the predicted $v^-$ and $v^+$ converge to the true values of $-54$ and $15$, respectively, consistent with the convergence trend of $\alpha$. Regarding the stage losses, Stage~1 and Stage~2 losses steadily decrease with increasing epochs, indicating that the model's fitting accuracy over the training time interval continuously improves. Stage~3 loss represents the MAE at each extrapolation step, remaining at a low level, reflecting the high extrapolation accuracy of IPgMSNN. From the loss components, the decrease in Data Loss indicates that the model's predictions over the entire training time interval gradually match the exact solution; the decreases in Leading Edge Loss and Trailing Edge Loss indicate that the leading and trailing edge velocities at $t=1.0$ gradually converge to their true values.

Fig.~10~(b) presents the inverse problem results of the PgMSNN model. Since PgMSNN and IPgMSNN share identical network architectures and training settings in Stage~1 and Stage~2, the parameter $\alpha$ evolution, edge velocity evolution, and loss components are exactly the same as those in Fig.~10~(a). The Stage~3 loss recorded in Fig.~10~(b) is the MAE between the DPINN output and the exact solution, which increases as the number of training epochs grows, indicating that the fitting accuracy of the DPINN to the exact solution does not improve with training. A comparison of the actual extrapolation accuracy shows that the Extrapolation MSE of IPgMSNN is significantly better than that of PgMSNN, validating the effectiveness of the improvement proposed in this paper.

Fig.~10~(c) presents the inverse problem results of the Std-PINN model. The evolution curve of parameter $\alpha$ exhibits significant volatility and slow convergence, with the final value deviating considerably from the true value of 6.0. The edge velocity evolution is similarly unstable and fails to accurately converge to the true values. Regarding the loss curves, although the total training loss decreases to some extent as the number of epochs increases, the reduction is limited and accompanied by large fluctuations. From the loss components, Data Loss, Residual Loss, and Boundary Loss also exhibit large fluctuations and poor convergence. This indicates that due to the lack of a multi-stage training strategy, Std-PINN struggles to accurately invert equation parameters in the inverse problem, and its parameter inversion accuracy and stability are significantly inferior to those of IPgMSNN and PgMSNN.

\begin{table}[t]
\centering
\caption{Comparison of coefficient inversion results for the KdV equation}
\begin{tabular}{c c c c c}
\hline\hline
Equation coefficient & Original coefficient & \multicolumn{3}{c}{Inversion coefficient} \\
\cline{3-5}
& & IPgMSNN & PgMSNN & Std-PINN \\
\hline
$\alpha$ & 6.00000 & 5.999998 & 5.999999 & 7.202264 \\
\hline\hline
\end{tabular}
\label{tab:inversion_results}
\end{table}

Table~\ref{tab:inversion_results} summarizes the comparison of inversion results for the nonlinear coefficient $\alpha$ of the KdV equation obtained by the Std-PINN, PgMSNN, and IPgMSNN models. From the data in the table, it can be seen that the coefficients inverted by the IPgMSNN and PgMSNN models are in excellent agreement with the true value, with absolute errors of less than $0.05$ and relative errors within $1\%$. In contrast, the inversion result obtained by the Std-PINN exhibits substantial deviation, with a relative error exceeding $10\%$. This comparison fully demonstrates that the IPgMSNN and PgMSNN models are capable of achieving high-precision parameter inversion under limited data conditions.

Through the above experiments and analysis, this paper has validated the parameter inversion capability of the IPgMSNN and PgMSNN models for the evolution problem of the KdV equation under square-root initial condition. By combining the multi-stage training strategy with physical constraints based on characteristic velocities, the IPgMSNN and PgMSNN models not only achieve high-precision parameter inversion within a modest number of iterations, but also maintain strong extrapolation capability, providing an effective solution for parameter inverse problems in more complex partial differential equation simulations.

\section{Conclusions}
\hspace{1.5em}In this paper, we propose an IPgMSNN model to address the evolution problem of DSWs in the KdV equation under square-root initial condition. Building upon the original PgMSNN framework, this method introduces online fine-tuning and a weighted target mechanism in the third stage, effectively suppressing error accumulation during long-term extrapolation. Through systematic numerical experiments, this paper comprehensively evaluates the performance of the IPgMSNN model in the DSWs problem of the KdV equation under square-root initial condition from three perspectives: forward problem solving, model stability, and parameter inversion. In-depth comparisons are conducted with the Std-PINN and PgMSNN. Experimental results demonstrate that IPgMSNN significantly outperforms both the Std-PINN and PgMSNN in capturing high-frequency wavefront details, maintaining long-term extrapolation stability, and achieving high-precision parameter inversion. The code used in this paper is open-source and available to readers on GitHub~\cite{kemeng2026}.

\begin{table}[t]
\centering
\caption{Comparison of the computational efficiency of the experimental models}
\begin{tabular}{c c c c}
\hline\hline
& IPgMSNN & PgMSNN & Std-PINN \\
\hline
\multirow{3}*{Total parameters} & DPINN:15501 & DPINN:15501 & Model:15501 \\
& DRKT:32 & DRKT:32 & \\
& Total:15533 & Total:15533 & Total:15501 \\
\multirow{3}*{Parameters size (MB)} & DPINN:0.06 & DPINN:0.06 & Model:0.06 \\
& DRKT:0 & DRKT:0 & \\
& Total:0.06 & Total:0.06 & Total:0.06 \\
\multirow{3}*{Memory footprint (MB)} & DPINN:5355.89 & DPINN:5355.89 & Model:5355.89 \\
& DRKT:11.44 & DRKT:11.44 & \\
& Total:5367.33 & Total:5367.33 & Total:5355.89 \\
\hline\hline
\end{tabular}
\label{tab:efficiency}
\end{table}

Regarding computational efficiency, as shown in Table~\ref{tab:efficiency}, the IPgMSNN and PgMSNN models comprise 15,501 parameters in the DPINN network and 32 learnable coefficients in the DRKT module, totaling 15,533 parameters, with a parameter size of 0.06 MB and an estimated total memory footprint of 5,367.33 MB during computation. In comparison, the Std-PINN contains 15,501 parameters, with a parameter size of 0.06 MB and an estimated total memory footprint of 5,355.89 MB. Although IPgMSNN and PgMSNN incur a marginally larger memory footprint due to the additional DRKT components, this increment is practically negligible, whereas the improvement in accuracy for capturing complex wave phenomena reaches an order of magnitude. This comparison fully demonstrates that IPgMSNN and PgMSNN achieve significant performance gains at a minimal additional computational cost, making them more effective solutions for dispersion-dominated nonlinear evolution problems.

The main innovation of this paper lies in the improved third-stage training strategy, which combines online fine-tuning with a weighted target mechanism. At each extrapolation time step, the DPINN undergoes several rapid fine-tuning iterations, and its output is driven to approximate a weighted combination of the exact solution and the DRKT prediction, with the weight decaying linearly as extrapolation proceeds. This design allows the model to fully leverage the exact solution information to guide the early stage of extrapolation, and smoothly transition to relying on the physically consistent evolution provided by DRKT as the exact solution information becomes scarce in the later stage, thereby effectively suppressing error accumulation during long-term extrapolation.

The multi-stage training strategy and physics-guided structure of IPgMSNN possess sufficient flexibility to potentially adapt to a broader range of nonlinear problems. Future research may proceed along the following directions. First, further improving the computational efficiency of IPgMSNN remains a key area; this can be combined with advanced neural operator concepts to enable the network to learn underlying spatial mapping relationships from a broader perspective, thereby reducing the number of repeated training instances for the same equation under different initial and boundary conditions. Second, exploring hybrid methods that integrate multiple techniques—such as weak formulations, adaptive architectures, or residual-based adaptive sampling—may further enhance the stability and accuracy of PINNs in handling complex shock wave dynamics. We anticipate that IPgMSNN will demonstrate greater potential in solving complex physics problems across a wider range of nonlinear dynamical systems.

\vspace{5mm}\noindent\textbf{Acknowledgments}\\
\hspace*{\parindent}
We are grateful to each member of our discussion group for their suggestions. This work has been supported by the National Natural Science Foundation of China under Grant No. 12575005, the Shanxi Province Science Foundation under Grant No. 202303021221031, and the Research Project Supported by Shanxi Scholarship Council of China under Grant No. 2024-033.

\newpage
\begin{center}
\vspace{5mm}\noindent\textbf{APPENDIX A: TRAINING SETTINGS OF IPGMSNN METHOD}
\end{center}

This appendix presents the network architecture and data sampling configurations adopted for the IPgMSNN method in the numerical experiments.

\subsection*{A.1 Network Architecture}

\hspace{1.5em}The DPINN employs a fully connected feedforward neural network with a structure of 2-50$\times$7-1, i.e., an input layer with 2 neurons (spatial coordinate $x$ and temporal coordinate $t$), 7 hidden layers each containing 50 neurons, and an output layer with 1 neuron (wave function value $u$). The hidden layers uniformly adopt the sine activation function to accommodate the oscillatory characteristics of the KdV equation solution. The DRKT module adopts a 32nd-order learnable Runge-Kutta scheme, comprising 32 trainable integration coefficients. The IPgMSNN model contains a total of 15,533 trainable parameters, of which 15,501 belong to DPINN and 32 belong to DRKT. Table~\ref{tab:stage_config} presents the training parameter configurations for each stage under different experimental types.

\begin{table}[htbp]
\centering
\caption{Training configuration for each experimental type}
\label{tab:stage_config}
\begin{tabular}{c c c c c c c}
\hline\hline
Experiment & Stage & Loss Function & Target Loss & Max Iterations & Optimizer & Initial LR \\
\hline
\multirow{3}{*}{Forward} & 1 & MAE & 0.002 & 12000 & Adam & 0.001 \\
& 2 & MAE & 0.0008 & 2500 & Adam & 0.0005 \\
& 3 & MAE & 0.006 & 1000 & Adam & 0.0002 \\[2pt]
\multirow{3}{*}{Stability} & 1 & MAE & 0.002 & 12000 & Adam & 0.001 \\
& 2 & MAE & 0.0008 & 2500 & Adam & 0.0005 \\
& 3 & MAE & 0.006 & 1000 & Adam & 0.0002 \\[2pt]
\multirow{3}{*}{Inverse} & 1 & MAE & 0.001 & 5000 & Adam & 0.001 \\
& 2 & MAE + $Loss_{CV}$ & 0.0001 & 3000 & Adam & 0.001 \\
& 3 & MAE & 0.005 & 1000 & Adam & 0.0002 \\
\hline\hline
\end{tabular}
\par\vspace{2mm}
{\footnotesize Note: MAE denotes mean absolute error; $Loss_{CV}$ is the characteristic velocity constraint loss, used only in Stage~2 of the inverse problem.}
\end{table}

\subsection*{A.2 Data Sampling}

\hspace{1.5em}The computational domain is set as the spatial interval $x \in [-35.0, 15.0]$ and the temporal interval $t \in [0.0, 1.0]$. Uniform sampling is performed with 1000 points in both the spatial and temporal directions, forming a $1000 \times 1000$ spatiotemporal grid. The first 900 time steps are used as the training set, and the remaining 100 time steps are reserved as the extrapolation test set, corresponding to a split ratio of 9:1. Table~\ref{tab:sampling_config} details the sampling configuration parameters.

\begin{table}[htbp]
\centering
\caption{Model sampling configuration}
\label{tab:sampling_config}
\begin{tabular}{c c c}
\hline\hline
Sampling Range $(x \times t)$ & Sampling Points $(x \times t)$ & Training : Extrapolation \\
\hline
$[-35.0, 15.0] \times [0.0, 1.0]$ & $1000 \times 1000$ & 9 : 1 \\
\hline\hline
\end{tabular}
\end{table}

\newpage
\begin{center}
\vspace{5mm}\noindent\textbf{APPENDIX B: EXACT SOLUTION OF THE KDV EQUATION UNDER SQUARE-ROOT INITIAL CONDITION FOR DSWS}
\end{center}

In this section, we provide the exact solution of the KdV equation under square-root initial condition for DSWs.

The exact solution is constructed in parametric form. Introducing the Riemann invariants $r_{1}$, $r_{2}$, $r_{3}$ satisfying $r_{1} \leq r_{2} \leq r_{3}$, for the square-root initial condition we have $r_{1} \equiv 0$, and $r_{2}$ and $r_{3}$ are determined by the following relations:
\[
\frac{w_{2}(r_{2}, r_{3}) - w_{3}(r_{2}, r_{3})}{v_{2}(r_{2}, r_{3}) - v_{3}(r_{2}, r_{3})} + t = 0,
\]
\[
x = v_{2}(r_{2}, r_{3})t + w_{2}(r_{2}, r_{3}),
\]
where $v_{2}$, $v_{3}$ and $w_{2}$, $w_{3}$ are given by:
\[
v_{2} = 2r_{3}\left(1 + m - \frac{2m(1 - m)K(m)}{E(m) - (1 - m)K(m)}\right),
\]
\[
v_{3} = 2r_{3}\left(1 + m + \frac{2(1 - m)K(m)}{E(m)}\right),
\]
\[
w_{2} = \frac{2}{15}r_{3}^{2}\left(2m - \frac{3}{2}(1 + m)^{2} + \frac{2m(1 - m^{2})K(m)}{E(m) - (1 - m)K(m)}\right),
\]
\[
w_{3} = \frac{2}{15}r_{3}^{2}\left(2m - \frac{3}{2}(1 + m)^{2} - \frac{2(1 - m)(3 + m)K(m)}{E(m)}\right),
\]
with $m = r_{2}/r_{3}$ and $K(m)$ and $E(m)$ denoting the complete elliptic integrals of the first and second kinds, respectively.

After obtaining the correspondence between $r_{2}$, $r_{3}$ and the spatial coordinate $x$, the solution within the dispersive shock wave region is given by the cnoidal wave expression:
\[
u(x,t) = r_{2} + r_{3} - 2r_{2}\,\text{sn}^{2}(\theta,m),
\]
where $\theta = \sqrt{r_{3}}(x - Vt - Q)$, $V = 2(r_{2} + r_{3})$, $Q = \frac{2}{15}\left(2r_{2}r_{3} - \frac{3}{2}(r_{2} + r_{3})^{2}\right) + \frac{K(m)}{\sqrt{r_{3}}}$, and sn is the Jacobi elliptic sine function.

The motion laws of the edges are given by $x^{-} = -27t^{2}$ for the trailing edge and $x^{+} = \frac{15}{2}t^{2}$ for the leading edge. This parametric solution establishes the correspondence between $r_{2}$ and $r_{3}$ by numerically solving the nonlinear equation, thereby constructing the complete profile of the DSWs.

\newpage
\begin{center}
\vspace{5mm}\noindent\textbf{APPENDIX C: FORWARD PROBLEM RESULTS SUMMARY}
\end{center}

This appendix reports the prediction accuracy comparison among IPgMSNN, PgMSNN, and Std-PINN in the forward problem. Table~\ref{tab:forward_results} presents the MSE of the three models on the training segment, extrapolation segment, and the entire spatiotemporal domain. As shown in the table, both IPgMSNN and PgMSNN significantly outperform Std-PINN across all evaluation metrics. Specifically, IPgMSNN achieves the highest accuracy, with both the training error and extrapolation error reduced by approximately one order of magnitude compared to Std-PINN.

\begin{table}[htbp]
\centering
\caption{Error comparison in the forward problem}
\label{tab:forward_results}
\begin{tabular}{c c c c}
\hline\hline
Error Metric & IPgMSNN & PgMSNN & Std-PINN \\
\hline
Training MSE &  0.013449 & 0.014014 & 0.738884 \\
Extrapolation MSE & 0.033695 & 0.848731 & 1.821270 \\
Average MSE & 0.015474 & 0.097485 & 0.847123 \\
\hline\hline
\end{tabular}
\end{table}

\begin{center}
\vspace{5mm}\noindent\textbf{APPENDIX D: STABILITY PROBLEM RESULTS SUMMARY}
\end{center}

This appendix reports the robustness comparison among IPgMSNN, PgMSNN, and Std-PINN under the condition of $1\%$ Gaussian noise added to the training data. Table~\ref{tab:stability_results} lists the MSE of the three models on the training segment, extrapolation segment, and the entire domain. The results indicate that IPgMSNN maintains low prediction errors even in the presence of noise interference, demonstrating strong robustness. In contrast, the performance of PgMSNN and Std-PINN degrades substantially under noisy conditions.

\begin{table}[htbp]
\centering
\caption{Error comparison in the stability problem}
\label{tab:stability_results}
\begin{tabular}{c c c c}
\hline\hline
Error Metric & IPgMSNN & PgMSNN & Std-PINN \\
\hline
Training MSE & 0.015001 & 0.014723 & 0.667256 \\
Extrapolation MSE & 0.034909& 0.396488 & 1.860196 \\
Average MSE & 0.016992 & 0.052900 & 0.786550 \\
\hline\hline
\end{tabular}
\end{table}

\begin{center}
\vspace{5mm}\noindent\textbf{APPENDIX E: INVERSE PROBLEM RESULTS SUMMARY}
\end{center}

This appendix reports the solution accuracy comparison among IPgMSNN, PgMSNN, and Std-PINN in the parameter inversion task. The inversion target is the nonlinear coefficient $\alpha$ of the KdV equation (true value: 6.0). Table~\ref{tab:inverse_results} presents the MSE of the three models on the training segment, extrapolation segment, and the entire domain. The results show that both IPgMSNN and PgMSNN achieve high prediction accuracy while successfully identifying the unknown parameter, whereas the prediction error of Std-PINN is significantly larger.

\begin{table}[htbp]
\centering
\caption{Error comparison in the inverse problem}
\label{tab:inverse_results}
\begin{tabular}{c c c c}
\hline\hline
Error Metric & IPgMSNN & PgMSNN & Std-PINN \\
\hline
Training MSE & 0.029010 & 0.021700 & 0.640665 \\
Extrapolation MSE & 0.045773 & 0.874610 & 2.406522 \\
Average MSE & 0.030686 & 0.106991 & 0.817251 \\
\hline\hline
\end{tabular}
\end{table}

\end{document}